\documentclass[appendixfloats,tighten,iop,numberedappendix]{emulateapj}
\usepackage[draft]{hyperref}
\usepackage{bm}
\usepackage{mathtools}

\begin{document}
\title{Methods of Reverberation mapping: I. Time-lag Determination by Measures of Randomness} 
\author{Doron Chelouche\altaffilmark{1,2}, Francisco Pozo Nu\~nez\altaffilmark{1,2},  and Shay Zucker\altaffilmark{3}}
\altaffiltext{1} {Department of Physics, Faculty of Natural Sciences, University of Haifa, Haifa 3498838, Israel; doron@sci.haifa.ac.il, francisco.pozon@gmail.com}
\altaffiltext{2} {Haifa Research Center for Theoretical Physics and Astrophysics, University of Haifa, Haifa 3498838, Israel}
\altaffiltext{3} {Department of Geosciences, Raymond and Beverly Sackler Faculty of Exact Sciences, Tel Aviv University, Tel Aviv 6997801, Israel; shayz@post.tau.ac.il}

\shortauthors{Chelouche, Pozo-Nu\~nez \& Zucker}
\shorttitle{Time-lag Determination by Measures of Randomness}

\begin{abstract}

A class of methods for measuring time delays between astronomical time series is introduced in the context of quasar reverberation mapping, which is based on measures of randomness or complexity of the data. Several distinct statistical estimators are considered that do not rely on polynomial interpolations of the light curves nor on their stochastic modeling, and do not require binning in correlation space. Methods based on von Neumann's mean-square successive-difference estimator are found to be superior to those using other estimators. An optimized von Neumann scheme is formulated, which better handles sparsely sampled data and outperforms current implementations of discrete correlation function methods. This scheme is applied to existing reverberation data of varying quality, and consistency with previously reported time delays is found. In particular, the size-luminosity relation of the broad-line region in quasars is recovered with a scatter comparable to that obtained by other works, yet with fewer assumptions made concerning the process underlying the variability. The proposed method for time-lag determination is particularly relevant for irregularly sampled time series, and in cases where the process underlying the variability cannot be adequately modeled.  
 
\end{abstract}

\keywords{
galaxies: active --- methods: data analysis --- methods: statistical --- quasars: general
}

\section{Introduction}

Reverberation mapping (RM) is a widely used technique for probing spatially unresolved regions in variable astronomical sources. Perhaps its best known application is for the study of the broad-line region (BLR) in quasars, where line emission from photoionized gas lags behind continuum variations, whence its spatial extent may be deduced \citep{bah72,che73}. In this context, RM is at the core of studies to estimate the mass of supermassive black holes \citep[e.g.,][]{kas00,pet00,pet04} with implications for the formation and coevolution of black holes and galaxies \citep[e.g.,][]{sh03,vol03,sch06,net07,tr12}. 

In its simplest form, RM seeks to determine a time delay between time series, which reflects on the size of the source that emits the lagging signal (be it in quasars or in X-ray binaries; \citealt{ut14}). Nevertheless, RM is not restricted to time-delay measurements, and additional information may be obtained on the geometry of the reverberating region by studying higher moments of the transfer function, provided the data are good enough \citep[e.g.,][]{bm82,kr91,pan12,gr13,poz13}. In what follows we restrict our discussion to time-delay measurement, which is a commonly encountered problem in astronomy. 

Astronomical techniques for RM are inherently different from radar techniques, for example, because the driving signal cannot be controlled and the sampling is rarely regular, which undermines many common algorithms for signal processing. Various approaches have been put forward to deal with irregular sampling. For example, linear interpolation of the data is commonly applied by various algorithms for time-delay determination that are based on cross-correlation techniques \citep[generally referred to in the literature as interpolated cross-correlation functions (ICCF), see, e.g.,][]{gas87,wel99,cd11,che13}. Other approaches that rely on more elaborate models include joint spline fitting of the quasar light curves with relative scaling and time-shifting, where the lag is obtained using a best-fit criterion \citep[and references therein]{te13,lia15}. Although these approaches to RM are quite effective, which has to do with the fact that quasar variability is characterized by soft power density spectra \citep{wp94}, they are not strictly justified because quasar light curves are stochastic in nature \citep{kel09}. 

With the advance in computing power, statistical models for quasar light curves \citep[e.g.,][]{s81,rp92} have been introduced to RM \citep[e.g.,][]{csr00,sug06,zu11}. These rely on the notion of quasar ergodicity, and often assume that the underlying process responsible for quasar variability is of the damped random-walk type. Whether this assumption is justified for all sources, at all times, and over the full range of variability timescales, is unclear \citep[but see also \citealt{f16}]{mu11,her15}.  

The aforementioned RM techniques, although widely used, require educated guessing as to how astronomical sources  behave when not observed, e.g. due to seasonal gaps, which might result in erroneous lags if not properly applied \citep{her15}. An alternative approach to RM relies on (largely) model-independent time-delay measurement techniques, which fall into two main categories: discrete cross-correlation function (DCF) schemes \citep[with its $z$-transformed variant, ZDCF, by \citealt{zdcf}]{ed88}, and serial-dependence schemes \citep[note that some of those have evolved beyond the notion of serial-dependence schemes and may rely on tunable parameters, e.g., \citealt{gm02} and references therein]{pel94,pel96}. While the use of discrete correlation schemes is widespread in reverberation studies \citep[e.g.,][]{col98,kas00,poz12}, they are often less stable when sparse sampling is concerned \citep{wp94,kas00}, and their results are sensitive to binning in correlation space \citep{kov14}. Serial-dependence schemes, on the other hand, have largely been restricted to lensing studies \citep[and references therein]{pel96,gm02,te13,lia15}, and their relevance to RM problems with nontrivial transfer function effects has not been adequately demonstrated, nor has their performance been benchmarked with respect to other commonly used RM schemes.

With the above limitations of current RM techniques in mind, searching for additional approaches to time-lag measurement is of the essence. Specifically, more reliable model-independent means for RM would help to refine current BLR-size--luminosity relations (e.g., by decreasing their scatter, \citealt{ha11}), fully exploit upcoming survey data \citep[e.g.,][]{sh15}, and assess the viability of quasars  as standard cosmological candles \citep{el02,cak07,wat11,cz13,ho17}. 

In this work we introduce a new class of techniques to quasar RM, which is associated with measures of randomness and includes measures of data-complexity and serial-dependence. Such techniques are extensively used in various non-RM contexts in other fields of science (e.g., in cryptography and econometrics) but are not as widespread in astronomy. Common to all methods employed here is the minimal set of assumptions made when extracting time delays. In particular, the methods outlined here do not rely on data interpolation, nor on binning in correlation space, and do not invoke arguments regarding quasar ergodicity. 

The paper is organized as follows: the method is described in \S2, where various statistical estimators are defined. Section 3 benchmarks several variants of the method using an extensive suite of numerical simulations. The methods are applied to RM datasets of varying quality in \S4, and their performance is  benchmarked against other methods of RM. The summary follows in \S5.

\section{Methods}
\label{alg}

\begin{figure}
\epsscale{1.18}
\plotone{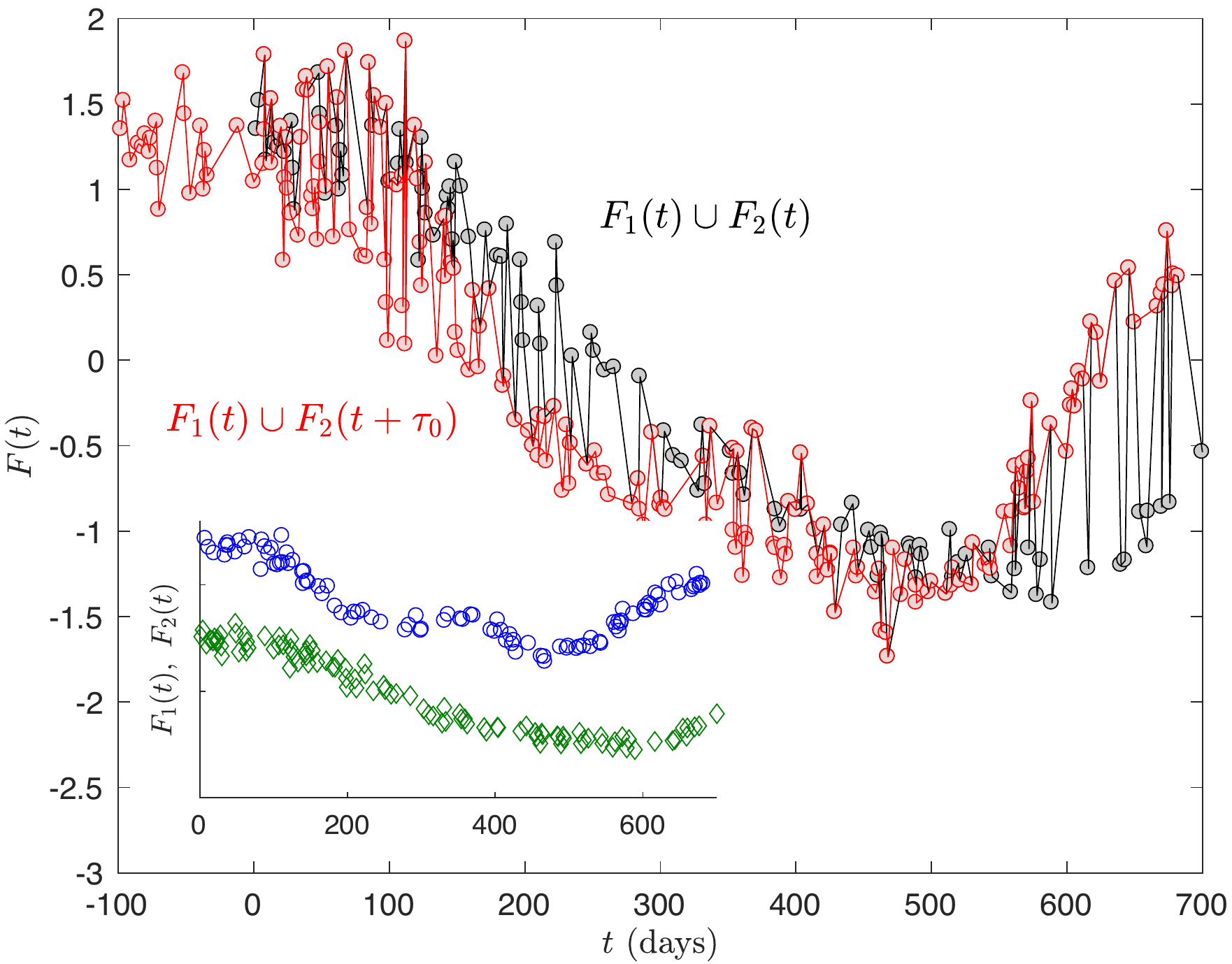}
\caption{The rationale behind an approach to reverberation mapping based on measures of randomness and serial-dependence. The inset shows two simulated light curves corresponding to the continuum signal (blue) and a line signal (green) that has been delayed and smeared by convolving with a wide kernel. Combining the two light curves into a single data stream, $F$, a higher level of regularity is obtained when a relative shift of $\tau=\tau_0=100$\,days is employed compared to that obtained by using other values of $\tau$ (see text). Statistical estimators that measure the level of regularity of $F$ can therefore be used to recover the time delay.}
\label{1}
\end{figure}

Let us represent two discrete light curves, $F_1$ and $F_2$, as two finite sets of ordered pairs:
\begin{equation}
F_1=\left \{ \left ( t_i^{(1)}, f_i^{(1)} \right ) \right \}_{i=1}^{N_1},~~~~F_2=\left \{ \left ( t_i^{(2)}, f_i^{(2)} \right ) \right \}_{i=1}^{N_2},
\end{equation}
where $f_i$ are the fluxes measured at times $t_i$. $N_1~(N_2)$ is the number of data points in $F_1~(F_2)$. We assign the indices of the ordered pairs so that they are in an increasing order of time stamps: $t_i^{(1)}<t_j^{(1)}$ when $i<j$ for $F_1$, and similarly for $F_2$. We further define the time-shifted version of $F_2$, 
\begin{equation}
F_2^\tau=\left \{ \left ( t_i^{(2)}+\tau, f_i^{(2)} \right ) \right \}_{i=1}^{N_2},
\end{equation}
where $\tau$ is a parameter.

Let us assume that $F_1$ and $F_2^\tau$ are normalized to zero mean and unit standard deviation (the degree to which this assumption is effective is discussed in detail in \S3.1), and that all time stamps are unique,\footnote{In the event that two data points have the same time stamp (this is generally unlikely, especially for $\tau\ne 0$), we introduce arbitrarily small shifts, which do not affect the end result, so that time stamps are unique.} i.e., $\forall i,j,\tau: t_i^{(1)} \ne t_j^{(2)}+\tau$. We then define the combined light curve as
\begin{equation}
F(t;\tau)=\left \{  \left ( t_i, f_i \right )  \right \}_{i=1}^N \equiv F_1 \cup F_2^\tau,
\label{fdef}
\end{equation}
where $N=N_1+N_2$. Here also we assign the indices in an increasing order of the time stamps, so that $F$ is an interlaced combination of $F_1$ and $F_2^\tau$. 

Clearly, $F(t)$ is a discrete function of time, which depends, by construction, on $\tau$. In particular, if $F_2$ is a version of $F_1$ that lags by some time interval $\tau_0$ -- as would be the case if $F_1/F_2$ is the continuum/broad-line light curve of quasars -- then $F$ will appear more regular\footnote{By the term "regular" we mean that longer trends in $F(t)$ become more pronounced and short term fluctuations are suppressed. For example, when $F_1$ and $F_2^\tau$ are, time-wise, better aligned, the power spectrum of $F$ becomes more reminiscent of red noise than of white noise.} when constructed from equation \ref{fdef} using $F_2^\tau$ with $\tau = \tau_0$ (Fig. \ref{1}). The problem of estimating the lag between $F_1$ and $F_2$ therefore translates to finding a reliable estimator for the regularity, or conversely the randomness of $F$, and defining a suitable extremum condition for it, from which the recovered time delay, $\tau_0^\ast$, may be deduced. Note that many of the statistical properties of $F$, such as its value-distribution moments, are independent of $\tau$ hence algorithms that are sensitive to the {\it ordering} of the data are required. 

\subsection{Definitions of Estimators}

There is vast literature on quantifying the serial-dependence, level-of-randomness, and complexity of datasets, which relates to coding theory (e.g., cryptography and data compression problems), and some algorithms work better than others, depending on the type of data and the purpose for which they are employed\footnote{A useful compendium is by \citet{nist}, which can be found at \url{http://csrc.nist.gov/publications/nistpubs/800-22-rev1a/SP800-22rev1a.pdf}.}. Below, we outline a few representative algorithms/statistics that operate on $F$ but we emphasize that additional relevant estimators exist, which are not included here and may be worth exploring. 

\subsubsection{Von Neumann}
\label{vn}

The mean-square successive difference after \citet{vn41} is 
\begin{equation}
\mathfrak{T}(\tau)=\frac{1}{N-1}\sum_{i=1}^{N-1}[F(t_i)-F(t_{i+1})]^2,
\label{vne}
\end{equation}
and recall the dependence of $F$ on $\tau$, namely Equation \ref{fdef}, whose explicit dependence is henceforth dropped for brevity. In particular, if $F(t_i)$ were to be drawn from a normal distribution then $\mathfrak{T}$ would converge to twice the variance for large $N$. If, however, positive correlations exist between successive points on the light curve (e.g., the long trends seen in quasar light curves) then its value would be smaller than twice the variance of the light curve, which is insensitive to the ordering of points. For the case in question, constructing $F(\tau)$ with $\tau\simeq \tau_0$, which corresponds to the lag between $F_1$ and $F_2$, is expected to result in pronounced trends in $F$ and hence to a minimum in $\mathfrak{T}(\tau)$ (Fig. \ref{2a}). The recovered lag, $\tau_0^\ast$, is defined here by the location of the minimum in $\mathfrak{T}$ over the $\tau$-range probed. A version of this test, which operates on the ranked light curve, $F_R(t)$, and leads to similar results, is due to \citet{bar82}. Similar results are also obtained when considering the sum of absolute differences rather than their squares (not shown). 

A version of the von Neumann (VN) estimator due to \citet[sometimes referred to as a "dispersion spectrum" or "Pelt statistics"]{pel94}, which is reminiscent of the $\chi^2$ definition, multiplies each of the terms in the above sum by a weighting factor, $W$, where
\begin{equation}
W_{i,i+1}=\frac{1}{\sigma^2(t_i)+\sigma^2(t_{i+1})}
\label{wflag}
\end{equation} 
($\sigma$ is the flux measurement uncertainty), and leads to a weighted version of the VN estimator. Like \citet{pel94}, we introduce a flag $G$, which multiplies each of the terms in the VN sum, yet our definition expands on theirs, and reads 
\begin{widetext}
\begin{equation}
G_{i,i+1}= \left \{
\begin{array}{ll}
\displaystyle 0 & \alpha_i+\alpha_{i+1}\ne 1~~\lor ~~ \alpha_i+\alpha_{i+1}=1 ~~\land ~~  \vert t_i-t_{i+1} +(\alpha_{i+1}-\alpha_i)\tau \vert \le \delta t  \\
\\
\displaystyle 1 & {\rm otherwise}
\end{array} \right . ,
\label{gflag}
\end{equation}
\end{widetext}
where $\alpha_i=1$ if the $i$'th data point originates from $F_2^\tau$ and $\alpha_i=0$ otherwise (see also Appendix A). This definition of $G$ serves for a dual purpose: it allows us to include in the sum only those cross terms of the light curve, where each point in the pair $\left ( F(t_i),F(t_{i+1}) \right )$ originates in a different light curve, thereby ignoring redundant information for time-delay measurements \citep{pel94}. Our definition of $G$ further allows us to exclude pairs of points whose {\it observed} difference in time-stamps is smaller than some timescale over which correlated errors may be important and could bias the results. For example, the effect of correlated errors between continuum and line flux measurements that originate from the same spectrum can be mitigated by using the $G$-flag with $\delta t=0$, as considered, for simplicity, henceforth (see \citealt{zdcf,al13}). The refined VN expression then takes the form 
\begin{equation}
 \mathfrak{T}(\tau)=\frac{\displaystyle \sum_{i=1}^{N-1}G_{i,i+1}W_{i,i+1}[F(t_i)-F(t_{i+1})]^2}{\displaystyle \sum_{i=1}^{N-1} G_{i,i+1} W_{i,i+1} }.
\label{vnn}
\end{equation}

\subsubsection{Kendall}
\label{secken}

This is, essentially, a non-parametric correlation estimator \citep{ken38} that involves the counting of the number of concordant and discordant pairs\footnote{Concordant pairs are those that are characterized by similar trends, i.e., the first point in both pairs having a smaller (larger) rank than the second point. Discordant pairs are those whose trends are opposite; see also \citet{zu15}.} ($N_c,~N_d$ respectively) in $F(t_i)$ or in its ranked version $F_R(t_i)$. It is defined as
\begin{equation}
\mathfrak{T}(\tau)=\frac{N_c-N_d}{(N-1)(N-2)/2},
\label{ken}
\end{equation}
and is larger for light curves that are more regular. In particular, when combining the light curves with the correct time shift (of order the lag), the estimator reaches a maximum at $\tau_0^\ast \simeq \tau_0$ (Fig. \ref{2a}). In addition to the standard implementation, we have also implemented a version of the test with the $G$-flag, as defined by equation \ref{gflag} with the appropriate normalization\footnote{The normalization term in the denominator of Eq. \ref{ken} then reads $\displaystyle \left ( \sum_i^{N-1}G_{i,i+1} \right ) \left ( (\sum_i^{N-1}G_{i,i+1})-1 \right )/2$.}.

\subsubsection{Wald--Wolfowitz}
\label{wwz}

\begin{figure}
\epsscale{1.18}
\plotone{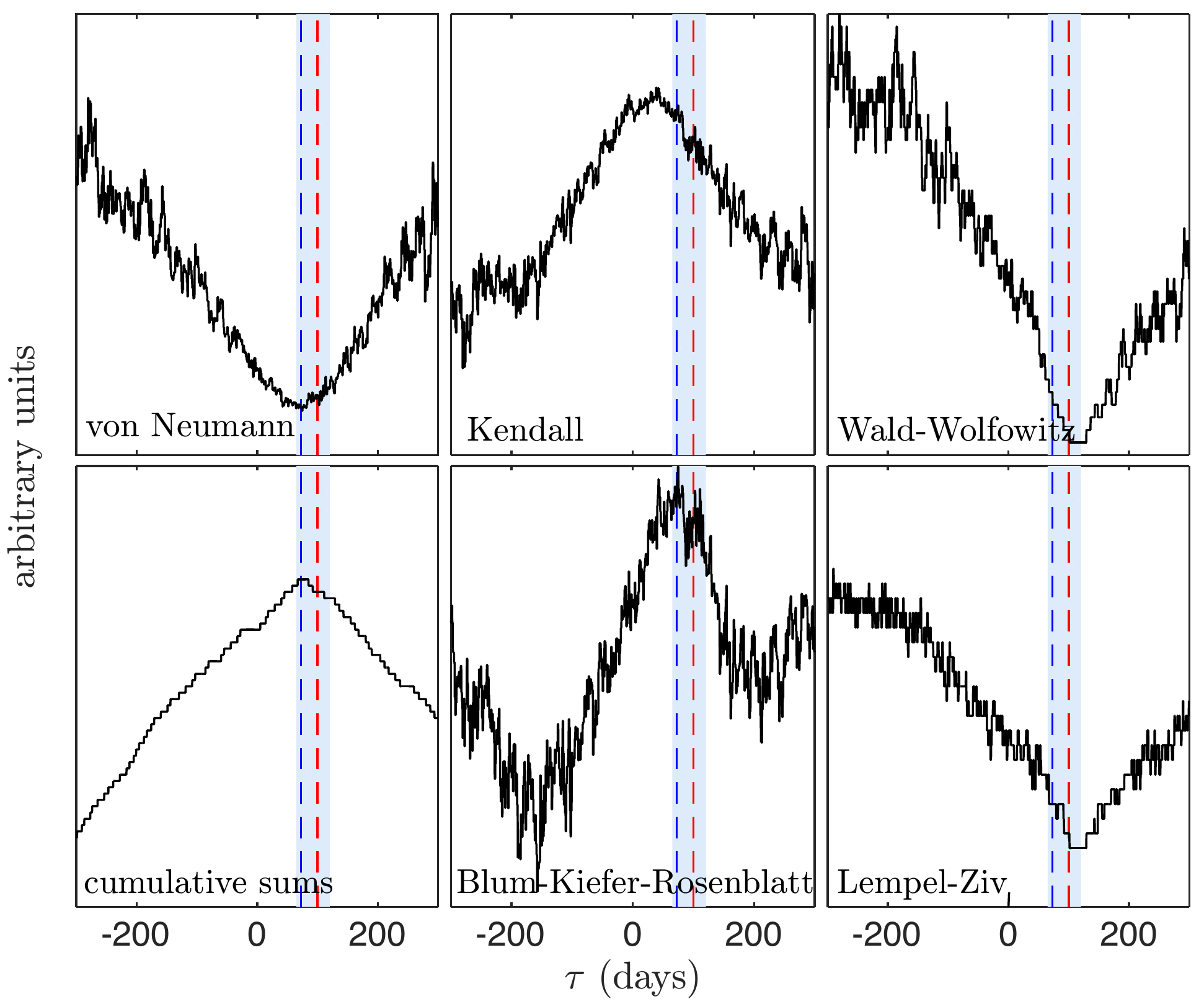}
\caption{Examples of measures of randomness of $F(t)$ as a function of $\tau$ (in all cases $F$ is comprised of the two specific light curves shown in the inset of Fig. \ref{1}). Clearly, all measures show significant dependence on $\tau$ and their extrema occur around the input lag of 100\,days (marked by the vertical dashed red line) hence $\tau_0^\ast\simeq \tau_0$. ZDCF results for the most likely lag, for the same particular set of light curves  shown in the inset of Figure \ref{1}, are shown by the dashed blue line, with asymmetric uncertainty intervals around it, based on a maximum likelihood approach \citep{al13}, drawn as blue shaded regions.}
\label{2a}
\end{figure}

Each point in the light curve is assigned with a property, $\mathcal{F}(t_i)$, which is unity if its value is above the median and zero otherwise. The following sum is then calculated \citep{ww40}:
\begin{equation}
\mathfrak{T}(\tau)=\sum_{i=1}^{N-1} \vert \mathcal{F}(t_i)-\mathcal{F}(t_{i+1}) \vert .
\label{ww}
\end{equation}
Light curves that show longer trends over many visits, and are hence less random, will result in smaller values of $\mathfrak{T}$. Conversely, more random light curves, or those that are non-random but rather exhibit frequent -- with respect to the cadence -- sawtooth-like behavior (not a characteristic of quasar light curves), will  attain higher values. We therefore identify the lag with $\tau$ that minimizes $\mathfrak{T}$ (Fig. \ref{2a}). In addition to the standard implementation, we have also implemented a version of the test with each term in the sum multiplied by the $G$-flag (Equation \ref{gflag}) with proper normalization maintained. Weighting by measurement uncertainties (Eq. \ref{wflag}) may also be formally included with proper normalization as in \S\ref{vn}.

\subsubsection{Cumulative sums}

This test assigns each point in $F(t_i)$ with a property $\mathcal{F}(t_i)$ so that  $\mathcal{F}(t_i)=1$ if $F(t_i)>\left < F \right >$ and $\mathcal{F}(t_i)=-1$ otherwise. If $\mathcal{F}(t_i)$ resembles a random walk then by summing over segments, one expects to observe relatively small deviations from zero (of the $\sqrt{N}$-type). On the other hand, larger deviations from a random-walk behavior are expected to occur  in light curves that include correlations and trends extending over many successive points or "steps" \citep[for additional information]{nist}. In particular, we calculate the maximum of partial sums,
\begin{equation}
\mathfrak{T}(\tau)= {\rm max} \left \{ S_n \right \}_{n=1}^N,~~~{\rm where}~~~S_n= \left \vert \sum_{i=1}^{n} \mathcal{F}(t_i) \right \vert.
\end{equation}
A maximal $\mathfrak{T}$ is expected to arise when $\tau$ is comparable to the time delay between $F_1$ and $F_2$ (Fig. \ref{2a}). As there is no clear dependence of the above sums on $N$ (quasar light curves are not of a random-walk type), we do not consider $W$-weights/$G$-flag implementations here.  

\subsubsection{Blum-Kiefer-Rosenblatt}

This test has evolved from the work of \citet{hf48}, and has been shown to have superior qualities in the astronomical context \citep{zu16}. The measure is defined using the ranked light curve as in \citet{bkr61}:
\begin{equation}
\mathfrak{T}(\tau)=\frac{1}{(N-1)^4}{\displaystyle \sum_{i=1}^{N-1} \left [ (N-1)c_i-F_R(t_i)F_R(t_{i+1}) \right ]^2},
\end{equation}
where $c_i$ is the "bivariate rank" \citep{zu16}, which is defined as the number of pairs of points $(F_R(t_{j}),F_R(t_{j+1}))$ for which $F_R(t_j)\le F_R(t_i)$ and $F_R(t_{j+1})\le F_R(t_{i+1})$. For light curves with longer trends (i.e., a lower level of randomness), the first term on the right hand side of the above expression would be larger. We therefore seek $\tau$ that maximizes $\mathfrak{T}$ and identify it with the lag (Fig. \ref{2a}). A $G$-flagged version has also been formally implemented with the appropriate normalization ($N-1 \to \sum_{i=1}^{N-1}G_{i,i+1}$). 

\subsubsection{Lempel-Ziv}

\citet{zip} introduced a complexity measure of a data stream, which is at the heart of a well known lossless data compression algorithm. A lower complexity measure means that the data are showing more significant trends and hence are easier to compress\footnote{For example, imagine $N$ points in the $(x,y)$ plane that trace a linear trend. These points may be characterized by $2N$ values or instead by $N+2$ values (their $x$-values augmented by the linear trend parameterization). Choosing the latter representation leads to a $\sim 50\%$ compression of the data volume without any information loss.}. We therefore expect $F(\tau)$ to have minimal complexity at around $\tau_0$ (Fig. \ref{2a}). To limit ourselves to a finite eigenvocabulary \citep{zip} we consider two binary-sequence options: either we operate on $\mathcal{F}(t_i)$, as defined in \S\ref{wwz}, or instead define  $\mathcal{F}(t_i)=1$ if $F(t_{i+1})-F(t_i)>0$ and $\mathcal{F}(t_i)=0$ otherwise ($\mathcal{F}$ is a sequence of $(N-1)$-points using this definition). It turns out that the outcome, as far as determination of time lag is concerned, depends little on the particular implementation used. Due to the binary nature of this estimator, and the uncertainties associated with its normalization, we do not consider error-weighting or $G$-flag implementations here.

\subsection{Definitions of Optimized Estimators}
\label{opt}

Consider each of the aforementioned estimators, which operate on a time-series $F$ defined in equation \ref{fdef}. Clearly, an improper relative normalization of $F_1$ and $F_2$ (or $F_2^\tau$) could lead to erroneous time-delays (see more below). Nevertheless, due to finite sampling, the true relative normalization of $F_1$ and $F_2$ is unknown. In this work we improve upon the method described in \citet{pel94}, and consider the following scheme when assembling an optimal version of $F$. Specifically, in analogy to equation \ref{fdef} let us define
\begin{equation}
F'(\tau, \eta , \epsilon)=F_1 \cup \left ( F_2^\tau+\eta F_2^\tau +\epsilon \right ),
\label{f1}
\end{equation}
where $\eta,~ \epsilon$ are parameters with $\eta> -1$ (unphysical normalizations of the light curves would be obtained for $\eta \le -1$ whereby $F_2^\tau$ is either nulled or its flux inverted). We note that $\vert \eta \vert \ll 1$ for light curves typical of RM and from symmetry considerations we may instead choose to normalize $F_1$ instead of $F_2^\tau$ so that an expression analogous to Equations \ref{fdef} and \ref{f1} would read
\begin{equation}
F''(\tau, \eta , \epsilon)=\left ( F_1-\eta F_2^\tau -\epsilon \right ) \cup F_2^\tau.
\label{f2}
\end{equation}
Clearly, the evaluation of $\mathfrak{T}$  should not depend on whether $F'$ or $F''$ is used as input, hence we define the optimized versions of the estimators as 
\begin{equation}
\mathfrak{T}(\tau,\eta,\epsilon)\equiv \frac{1}{2}(\mathfrak{T}'+\mathfrak{T}''),
\label{dd0}
\end{equation} 
where $\mathfrak{T}',~\mathfrak{T}''$ are calculated for $F',~F''$ respectively. An extremum can then be sought either numerically or analytically in the three-dimensional phase space spanned by $\eta,~\epsilon$ and $\tau$ (see Appendix A for the VN scheme).  The aforementioned symmetrization is robust to situations where little overlap exists between the shifted light curves,  e.g., due to seasonal gaps. In particular, an asymmetric implementation of the normalization scheme may lead to either $\eta \to 0$ (in the case of a minimum) or $\eta \to \infty$ (in the case of a maximum) solutions, and thence to erroneous time-delay measurements.

\section{Simulations}

\begin{figure}
\epsscale{1.16}
\plotone{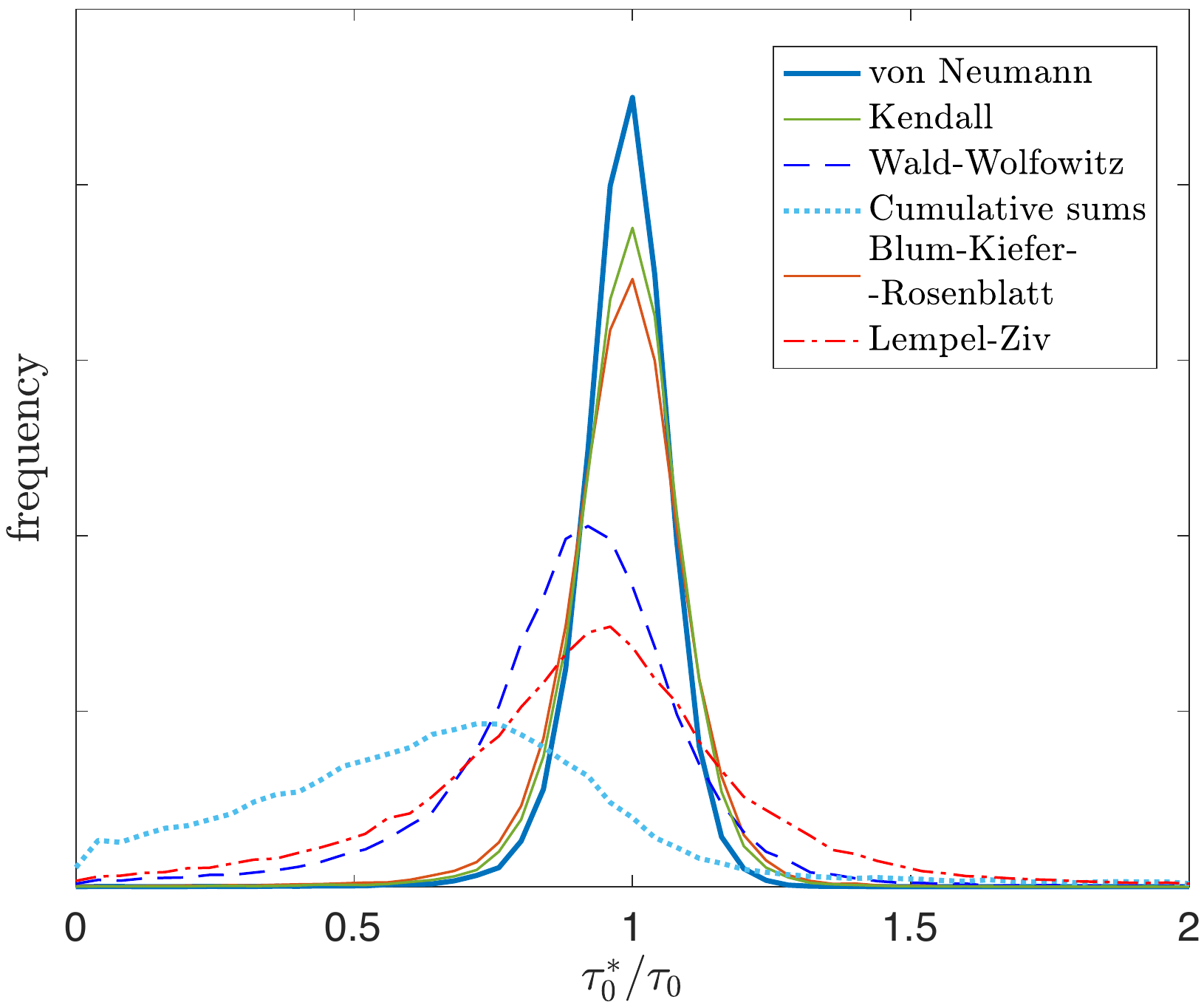}
\caption{Recovered distributions of time-delay ($\tau_0^\ast$) for the various measures of randomness considered in this work (see legend). Evidently, the VN estimator is sharply peaked at the input lag and is characterized by the smallest dispersion around this value. Other estimators, especially Lempel--Ziv, cumulative sums, and Wald--Wolfowitz (non-solid lines) are less competitive for time-lag determination and may even be biased (see text).}
\label{2b}
\end{figure}

To test the efficiency of measures of randomness for time-lag determination, and to identify potential biases with respect to the true lag, we resort to simulations. We model the driving continuum light curve, $F_1(t)$, using the method of \citet{tm95} with the Fourier transform $\tilde{F_1}(\omega)\propto  \omega^\gamma$, where $\gamma \lesssim-1$, which is typical of quasars over days to years timescales \citep{giv99,mu11,cap16}\footnote{Note that some works define the powerlaw index of the power spectral density instead.}. 

The light echo is obtained from the convolution of the driving light curve with a transfer function $\psi$, so that $F_2= F_1 * \psi$, where the input lag, $\tau_0 =\int t \psi(t)$, and $"*"$ denotes convolution. In this work we consider rectangular transfer functions over the time range $t\in [\tau_0-\Delta \tau/2,\tau_0+\Delta \tau/2]$ where $\Delta \tau/\tau_0 \le 2$ thus preserving causality while allowing us to control the level of BLR isotropy around the ionizing source by the ratio $\Delta \tau/\tau_0$ \citep[e.g.,][and references therein]{poz13}.

To simulate light curves that are reminiscent of astronomical data, $F_1$ and $F_2$ are {\it independently} and {\it randomly} sampled with $N_1=N_2=N/2$ visits. For comparison, we also experimented with simultaneous uniform sampling of both time series and found the relative benchmarks of the various estimators to be qualitatively similar, hence they are not shown.  Lastly, we add uncorrelated (Gaussian) noise to the sampled light curves to mimic observational uncertainties. 

\begin{figure*}
\epsscale{1.15}
\plottwo{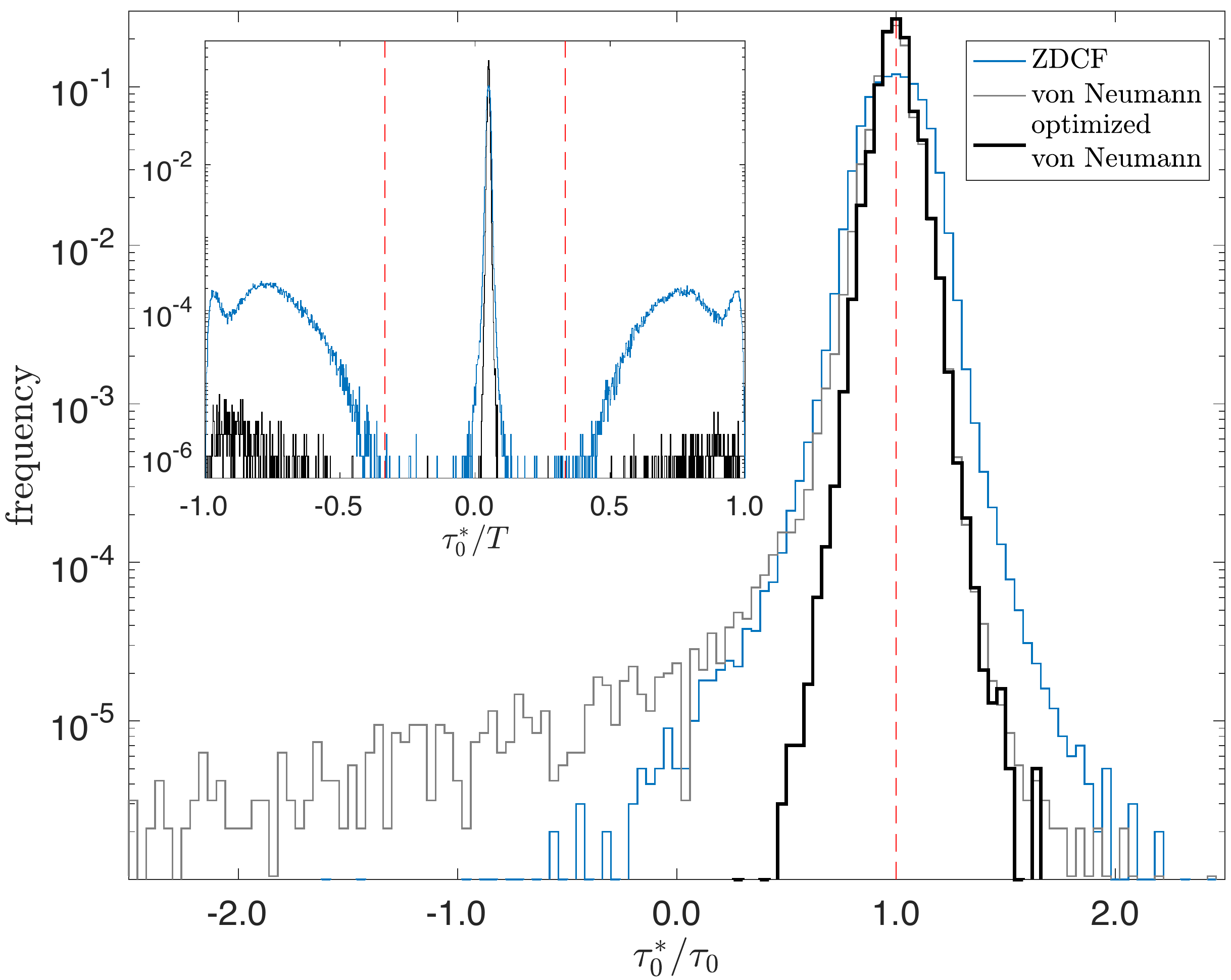}{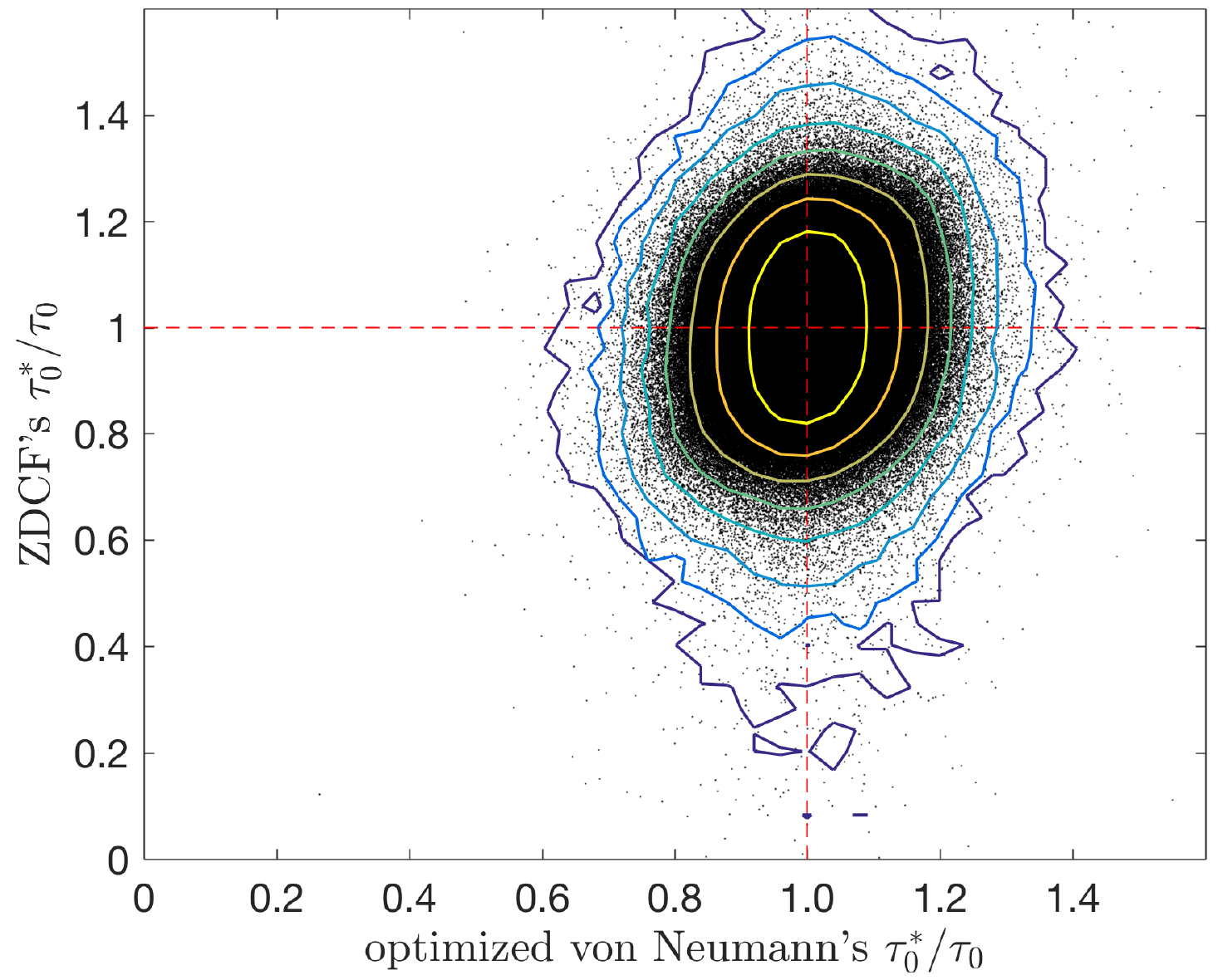}
\caption{Comparison of the VN and ZDCF schemes for time-delay measurement using mock light curves. {\it Left:} the distributions of $\tau_0^\ast$ based on the VN estimator (gray histogram), the optimized VN estimator (black histogram, see \S\ref{opt}), and the ZDCF estimator (blue histogram); note the logarithmic scaling. Both versions of the VN estimator outperform the ZDCF with higher peak values and narrower distribution cores. The optimized VN scheme results in significantly lower rms deviations from $\tau_0^\ast/\tau_0=1$ (marked by the dashed red line). The inset shows the full range of $\tau_0^\ast/T$ obtained for the ZDCF and for the optimized VN (note the small-number statistics in the lower part of the diagram). The large suppression of spurious lags with $\vert \tau_0^\ast/T \vert \gtrsim 0.3$ using the VN scheme compared to the ZDCF method is noticeable (see text). {\it Right:} The optimized VN vs. the ZDCF time delays in units of the input lag, $\tau_0$ for $\sim 10^6$ sets of mock light curves. Note the lack of correlation between the bulk of the results of the two methods (see text), and the narrower distribution obtained for the VN scheme.}
\label{comparison}
\end{figure*}

Having quasar RM campaigns in mind, our basic set of light-curve realizations, around which parameterized excursions are later explored (\S3.1), is defined by $\gamma=-1$ \citep{kel09}, $\tau_0=100$\,days \citep{kas00}, and $T/\tau_0=20$, which is the ratio of the total span of the time series, $T$, to the lag (see Fig. 8 in \citealt{wel99} for motivation). In addition, each light curve is randomly sampled with 100 visits ($N=200$), which corresponds to an average sampling timescale for each light curve of $\left < t_{\rm sampling} \right> \simeq 0.2\tau_0$ \citep{poz12,bar15,du16}. We further take $\Delta \tau/\tau_0=0.5$, which is consistent with a relatively face-on disk geometry \citep{pan12,poz13}. A signal-to-noise ratio S/N$=33$ is assumed, which corresponds to measurement uncertainties at the $\sim 3\%$ level \citep[see][]{kas00}, and may be compared to the assumed fractional rms  amplitude of variability of the quasar, $F_{\rm var}=0.2$ \citep{rod97,kas00}. 

The time-delays between pairs of mock light curves are deduced in a fully automated way by searching for the  extrema associated with each of the aforementioned measures of randomness. Unless otherwise specified, we do not use an error-weighting scheme, nor $G$-flag implementations. 

The recovered distributions of time-lag, $\tau_0^\ast$, are based on $\gtrsim 10^4$ light-curve realizations, and are shown for the non-optimized estimator versions in Figure \ref{2b}. We find that the Bartels and VN estimators lead to similar results, and only the latter, which leads to somewhat improved lag statistics, is shown. Further, the VN estimator is the best performer in our simulations: it exhibits the narrowest distribution of $\tau_0^\ast$, and is relatively unbiased with $\left < \tau_0^\ast \right > \simeq \tau_0$. Next in terms of estimator performance come the Kendall and Blum--Kiefer--Rosenblatt estimators, which exhibit larger dispersions around the mean. The Wald--Wolfowitz, the Lempel--Ziv -- the particular implementation of which appears to be largely immaterial -- and the cumulative sums estimators are, in order of appearance, the worst performers: they lead to the broadest distributions around the central value, and their means, like their median and centroid values, and are biased to shorter lags -- by up to 30\% for the cumulative sums estimator -- for our set of simulations.

\subsection{Benchmarking the VN Estimator}

With the above results in mind, we focus in this section on benchmarking and improving the von Neumann scheme for RM purposes, and compare its deduced lags to those obtained by the ZDCF method of \citet{zdcf,al13},\footnote{See the complete code package at: \url{http://wwo.weizmann.ac.il/weizsites/tal/research/software/}.} which is a widely used discrete correlation scheme in the field of RM. Unless otherwise specified, the ZDCF algorithm is used here in its default optimal-binning mode and with zero-lag points omitted \citep{al13}. 

A detailed comparison of the VN and ZDCF deduced lags for $10^6$ mock light curves defined by our basic parameterization is shown in Figure \ref{comparison}. We find that the von Neumann estimator provides a distribution of laga with a narrower core than the ZDCF -- the ratio of their full widths at half-maximum is $\sim 0.7$ for the simulated case; see Fig. \ref{comparison} -- but with a more pronounced long-tail wing extending to short and even negative time delays (the latter are obtained in $\sim 0.05\%$ of the realizations for this set of simulations). Detrending of the light curves appears to be counter-productive for the VN approach and hence not is considered in the remainder of the paper. We attribute the emergence of the wing toward short time delays to relative  inconsistencies in normalization in a fraction of the light-curve realizations, and demonstrate the improved statistics obtained for the optimized scheme of \S\ref{opt} in Figure \ref{comparison}. In particular, the distribution of $\tau_0^\ast$ obtained by the optimized scheme becomes more symmetric and narrow; it is significantly ($\sim 40$\%) narrower than the ZDCF, with the former having 68\% of the $\tau_0^\ast$ values within $\pm0.35 \left < t_{\rm sampling} \right > \simeq \pm 7$\,days from the peak for our basic set of simulations. 

The optimized scheme appears to be considerably better than the ZDCF in suppressing spurious time delays over the range $[-T,T]$ (see the inset of Fig. \ref{comparison}). Nevertheless, in what follows we restrict our time-delay search window to the range $[-T/3,T/3]$, within which more than half of the light-curve features covered by the time series are in the time-wise overlapping region between $F_1$ and $F_2^\tau$ (this statement is true in the statistical sense but may not hold for individual light curve realizations).

We further note the overall poor correspondence between the time-delay measurements of the VN and the ZDCF methods, leading to a Pearson correlation coefficient $\rho <0.01$ (righthand plot of Fig. \ref{comparison}). Recovered time delays that deviate from the central value by more than $\left < t_{\rm sampling} \right >$ in either direction, for both methods, are somewhat better correlated (note the somewhat tilted outer contours in Fig. \ref{comparison} characterized by $\rho\lesssim 0.5$).  Interestingly, the poor correlation between the time delays obtained by the two methods implies that they may be viewed as independent measurements of the lag, and under certain circumstances might even be used in tandem to improve time-delay estimates; a proper treatment of this possibility is beyond the scope of the present work.

\begin{figure}
\epsscale{1.22}
\plotone{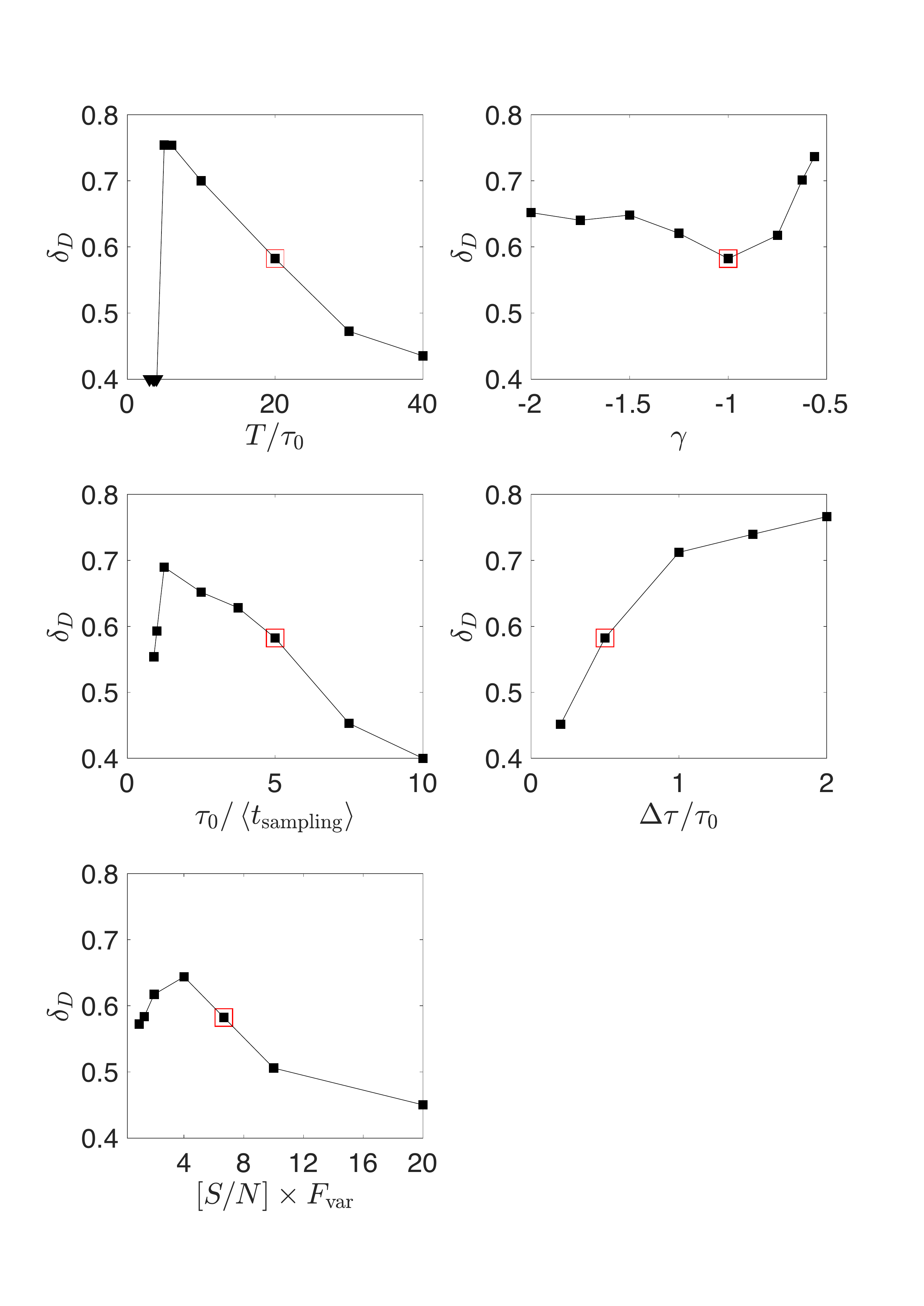}
\caption{The relative performance of the optimized VN and the ZDCF methods ($\delta_D<1$ implies an advantage to the VN method; see text). Each panel shows an excursion from the basic model, whose results are highlighted in red, by varying one parameter at a time (left panels relate to observational parameters and right panels to quasar parameters). For clarity, triangles mark upper limits, with the actual values being $\sim 0.2$. Generally, the performance of the optimized VN estimator exceeds that of the ZDCF for the parameter range explored here.}
\label{5}
\end{figure}

We next benchmark the performance of the optimized VN method against the ZDCF for various parameterized excursions from our basic set of light-curve realizations. To this end we define the rms deviation,\footnote{The root mean square deviation is defined here as $\displaystyle D=\sqrt{N_s^{-1}\sum_{i=1}^{N_s} (\tau_{0,i}^\ast-\tau_0)^2}$ where $N_s\ggg1$ is the number of realizations. To avoid having our statistics influenced by $\tau_0^\ast$-outliers (see Fig. \ref{comparison}), $D$ includes the sum of all but the extreme 2.5\% measurements on either side of the $\tau_0^\ast$-distribution, for either of the estimators.}, $D$, of the measured lag from the true lag per estimator, and calculate the ratio $\delta_D\equiv D_{\rm von~Neumann}/D_{\rm ZDCF}$ ($\delta_D<1$ implies an advantage to the VN approach). The results are summarized in Figure \ref{5}, where each point is based on  the analysis of $10^5$ mock light curves.

We find the performance of the optimized VN scheme to be consistently better than that of the ZDCF method over much of the parameter space explored here. Specifically, it performs best for slopes, $\gamma\simeq -1$, of the order of the observed values (but underperforms for $\gamma>-0.5$, not shown) and for narrow transfer functions, such as those that characterize relatively face-on disks or rings \citep{pan12}. A further advantage of the optimized VN scheme is attained for more extended campaigns (with a fixed number of points), mainly because of the increased scatter in the ZDCF results with poorer sampling. Higher cadence and S/N further improve the relative performance of the optimized VN scheme, as may be expected from a robust statistical estimator given the rising quality of the data. Interestingly, considering a model with S/N$\times F_{\rm var}=20$, $T/\tau_0=40$, and $\tau_0/t_{\rm sampling}=10$, and repeating the simulations, results in the distribution of $\tau_0^\ast$ of the optimized VN scheme being $\sim 10^2$ times narrower than the ZDCF (not shown).

As data quality deteriorates (e.g., sparser light curves), the deduced distributions of $\tau_0^\ast$ broaden and may blend with the spurious lag population at $\vert \tau_0^\ast \vert \gtrsim T/3$ which is shown in the inset of Figure \ref{comparison}. Nevertheless, our calculations indicate a consistent advantage of the VN scheme over the ZDCF approach also in this limit. This is most notable for sparse sampling with $\left < t_{\rm sampling} \right > \gtrsim \tau_0$ and when the total number of points per light curve is low  ($\gtrsim 18$ for this set of simulations; enforcing a different ZDCF binning scheme than the default does not appreciably change the results and hence is not shown). A significant relative decline in ZDCF's performance is seen also for the short simulated campaigns with $T/\tau_0< 5$, which rarely characterizes RM campaigns because of well known biases \citep{wel99}. In the example considered here, the ZDCF time-lag distributions develop a non-negligible wing extending to short/negative delays while the optimized VN is not substantially affected and hence is less biased. A more in-depth exploration of the relative performance of the two estimators under challenging observing conditions, which are not conducive to reliable RM \citep{wel99}, is beyond the scope of this paper.

Finally, we note that a similar behavior characterizes also the Bartels estimator, whose results are not shown here. Specifically, its $\delta_D$ values are generally smaller than unity, and hence its performance is superior to the ZDCF, but they are $\lesssim10\%$ larger than VN's over the parameter range considered here.

\subsubsection{The Weighted VN Scheme}
\label{flags}

Here we consider the effect of spikes in the light curve (due to larger measurement uncertainties in a fraction of the data points) and correlated noise between the light curves on the deduced time delays and the means to mitigate them using the weighted VN scheme of \S\ref{vn}. To this end, more realistic sets of light curves have been simulated, which mimic typical light curves used for (a) BLR RM and (b) continuum time-delay measurements (perhaps associated with the accretion disk; \citealt{col98}). 

\begin{figure}
\epsscale{1.17}
\plotone{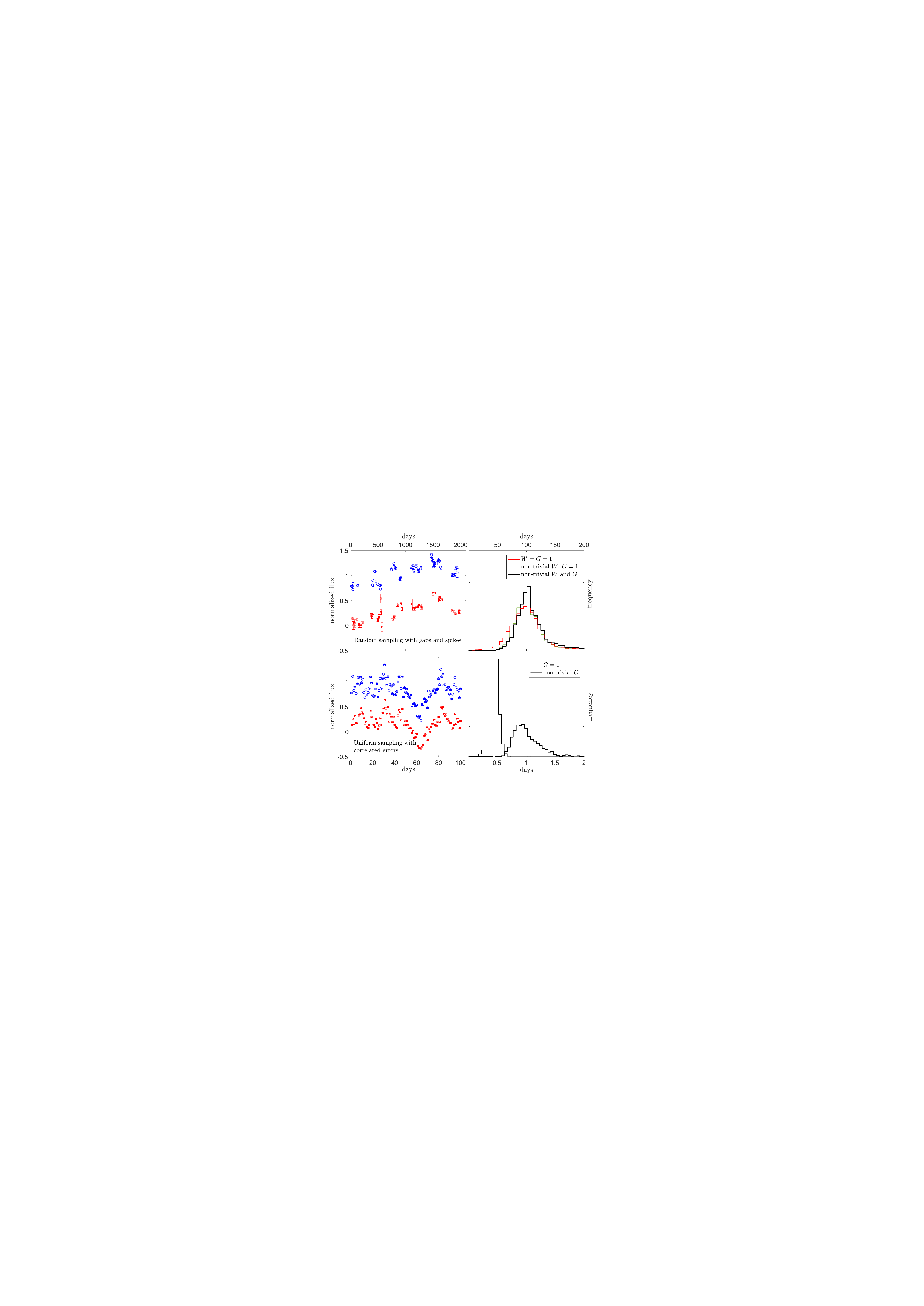}
\caption{Application of the weighted optimized VN scheme to randomly sampled quasar light curves with gaps and spikes (upper panels) and to uniformly sampled light curves with correlated errors and uniform measurement uncertainties (lower panels). Recovered time-lag distributions for $10^4$ realizations (with FR/RSS implementation; see \S4) are shown in the right panels while the left panels show typical examples for continuum (blue) and emission line (red) mock light curves. The different distributions shown pertain to different implementations of the algorithm (see legend and text). }
\label{weights}
\end{figure}

To better mimic BLR RM data we assume our basic set of simulations but include seasonal gaps 150\,days wide, and assume that 10\% of the data points suffer from an enhanced noise level, with their measurement uncertainties being larger by a factor of 3; an example is shown in Figure \ref{weights}. Results for the recovered time delays from $10^4$ mock light curve simulations using the weighted and unweighted VN schemes are shown in Figure \ref{weights}, and demonstrate the advantage of using the former. We find marginal advantage in using the $G$-flag scheme. 

To study the effect of correlated noise we turn to the case of continuum RM, where relevant light curves are often characterized by high, nearly uniform cadence, and we rely also on spectroscopic data for measuring fluxes at different wavelengths, hence leading to correlated errors. In our set of simulations we assume a time delay $\tau_0=1$\,day with $\Delta \tau/\tau_0=2$. Sampling cadence is fixed at 1\,day, hence the transfer function is barely resolved. We further consider correlated noise at the 5\% level, which exceeds the 1\% formal measurement uncertainties assumed here, and depict one such realization in Figure \ref{weights}. Results for the weighted VN give a distribution of $\tau_0^\ast$ centered\footnote{The fact that time-delays on subsampling timescales are obtained should not come as a surprise as those reflect on the {\it average} numbers obtained by the FR/RSS scheme (see \S4).} around 0.5\,days, which does not extend to values as large as the input delay because of correlated errors. The bias is alleviated using our $G$-flag implementation, and the average delay is consistent with the input lag. 

\section{Data Applications}

In this section we apply the schemes to measure randomness (\S2) to quasar data of varying quality, and benchmark their performance. We aim to provide lag measurements for individual sources and epochs, so time-lag uncertainties need to be quantified. Here we resort to the standard techniques in RM for estimating uncertainties in lags \citep[see][for a different implementation]{pel94}, which involves the FR/RSS algorithm of \citet{pet98,pet04}, and quote upper/lower uncertainties that correspond to 16/84 percentiles as obtained using the lag statistics for $10^4$ realizations for each set of light curves (see Appendix B). Unless otherwise stated, optimized and weighted versions of the estimators (\S2.2), including nontrivial $G$-factor implementations, are considered where applicable.

\begin{figure}
\epsscale{1.2}
\plotone{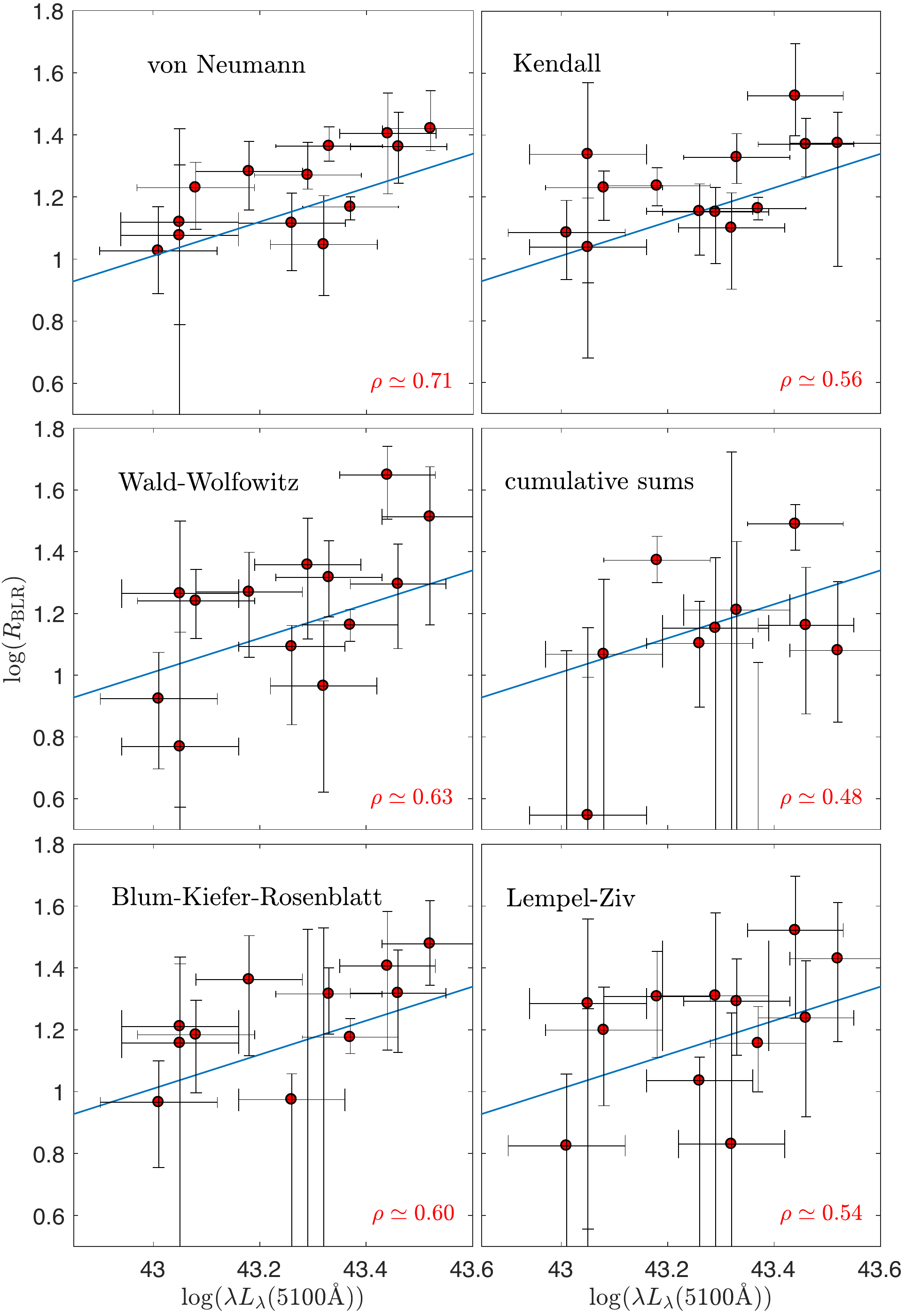}
\caption{BLR-size--luminosity relation for multi-epoch data for NGC\,5548. Data were taken from Peterson et al. (2002) with host-corrected luminosity estimates from \citet{ben13}. The blue line is the best-fit relation of Bentz et al. (2013). Measurement uncertainties were obtained using the FR/RSS scheme of \citet{pet04}. The Pearson correlation coefficient, $\rho$, is denoted in each panel, and implies an advantage for the VN measure over the other methods described here.}
\label{N5548}
\end{figure}

\begin{table}
\caption{Von Neumann Analysis of the Sample of Active Galactic Nuclei of \citet{ben13}}
\label{T1}
\vspace{-0.22in}
\begin{center}
\begin{tabular}{llllll}
	    & 	    	&	& $\tau_{\rm VN}$	& $\tau_{\rm Bentz}$	 & \\
object &  $z$	& 	 ${\rm log}(L_{\rm opt})$ 		& (days$^\dagger$)		& (days$^\dagger$) & Ref.\\
\tableline

Mrk\,335	&0.026	& $43.68\pm0.06$	& $13.7^{+0.9}_{-1.3}$	& $14.3\pm0.7$ & 1 \\

PG0026	&0.142	& $44.91\pm0.02$	& $79^{+32}_{-24}$	& $111.0^{+24.1}_{-28.3}$ & 2 \\

PG0052	&0.155	& $44.75\pm0.03$	& $68\pm20$	&$89.8^{+24.5}_{-24.1}$ & 2 \\

Fairall 9	&0.047	& $43.92\pm0.05$ &  $34^{+15}_{-25}$	&  $17.4^{+3.2}_{-4.3}$ & 3 \\

Mrk\,590	&0.026	& $43.53\pm0.07$ 	& $15.8^{+6.4}_{-9.1}$	&  $20.7^{+3.5}_{-2.7}$ & 4 \\

3C\,120	&0.033	& $43.87\pm0.05$	& $26.7^{+2.1}_{-1.7}$	&  $25.9\pm2.3$	&  1 \\

Ark\,120	&0.033	& $43.92\pm0.06$	& $34.9^{+12.6}_{-13.6}$	&  $47.1^{+8.3}_{-12.4}$ & 4 \\

Mrk\,79	&0.022	& $43.57\pm0.07$ 	& $25^{+11}_{-10}$ 	&  $16.0^{+6.4}_{-5.8}$ & 4 \\

PG0804	&0.100	& $44.85\pm0.02$		& $123^{+25}_{-37}$	& $146.9^{+18.8}_{-18.9}$ & 2 \\

Mrk\,110	&0.035	& $43.59\pm0.11$	& $23.7^{+8.6}_{-7.8}$	&  $26\pm7$ & 4 \\

PG0953	&0.234	& $45.13\pm0.01$	& $154^{+65}_{-29}$	&  $150.1^{+21.6}_{-22.6}$ & 2 \\

NGC\,3227	&0.004	& $42.24\pm0.11$ 	& $2.9^{+2.6}_{-2.1}$	&  $3.75^{+0.76}_{-0.82}$ & 5 \\

Mrk\,142	&0.045	& $43.54\pm0.04$ 	& $3.2^{+2.4}_{-5.3}$	&  $2.74^{+0.73}_{-0.83}$ & 6 \\

NGC\,3516	&0.009	& $42.73\pm0.21$ 	& $8.8^{+2.2}_{-1.1}$	&  $11.68^{+1.02}_{-1.53}$ & 5 \\

SBS\,1116 	&0.028	& $42.07\pm0.28$ 	& $1.2\pm 1.2$	&  $2.31^{+0.62}_{-0.49}$ & 6 \\

Arp\,151 	&0.021	& $42.48\pm0.11$ 	& $3.0^{+1.0}_{-0.9}$	&  $3.99^{+0.49}_{-0.68}$ & 6 \\

NGC\,3783	&0.010	& $42.55\pm0.18$ 	& $5.2^{+5.1}_{-4.8}$	&  $10.2^{+3.3}_{-2.3}$ & 7 \\

Mrk\,1310	&	0.020	& $42.23\pm0.17$ 	& $2.6^{+1.8}_{-1.2}$	&  $3.66^{+0.59}_{-0.61}$ & 6 \\

NGC\,4051	&0.002	& $41.96\pm0.20$ 	& $1.5^{+1.1}_{-1.3}$	&  $1.87^{+0.54}_{-0.50}$ & 8 \\

NGC\,4151	&0.003	& $42.09\pm0.22$ 	& $7.1^{+1.7}_{-1.9}$ &  $6.58^{+1.12}_{-0.76}$ & 9 \\

Mrk\,202 &0.021	& $42.20\pm0.18$ & $2.1^{+3.0}_{-2.0}$	&  $3.05^{+1.73}_{-1.12}$ & 6 \\

NGC\,4253	&0.013	& $42.51\pm0.13$ 	& $4.0^{+1.0}_{-1.1}$	&  $6.16^{+1.63}_{-1.22}$ & 6 \\

PG1226	&0.158	& $45.90\pm0.02$ 	& $340^{+127}_{-123}$	&  $306.8^{+68.5}_{-90.9}$ & 2 \\

PG1229	&0.063	& $43.64\pm0.06$ 	& $30^{+30}_{-17}$	&  $37.8^{+27.6}_{-15.3}$ & 2 \\

NGC\,4593	&0.009	& $42.87\pm0.18$ 	& $3.6^{+1.8}_{-3.2}$	&  $3.73\pm0.75$ & 10 \\

NGC\,4748	&0.015	& $42.49\pm0.13$ 	& $5.6^{+4.6}_{-4.4}$	&  $5.55^{+1.62}_{-2.22}$ & 6 \\

PG1307	&0.155	& $44.79\pm0.02$ 	& $89^{+72}_{-74}$	&  $105.6^{+36.0}_{-46.6}$ & 2 \\

Mrk\,279	&0.031	& $43.64\pm0.08$ 	& $16.1^{+3.4}_{-3.8}$	&  $16.7\pm3.9$ & 11 \\

PG1411	&0.090	& $44.50\pm0.02$ 	& $98^{+79}_{-127}$ &  $124.3^{+61.0}_{-61.7}$ & 2 \\

NGC\,5548	&0.017	& $42.93\pm0.12$ 	& $16.1^{+4.4}_{-3.6}$	&  $12.40^{+2.74}_{-3.85}$ & 5 \\

	&	& $43.33\pm0.10$ 	& $22.6^{+3.4}_{-2.5}$	&  $19.7\pm1.5$ &  12 \\
	&	& $43.08\pm0.11$ 	& $16.7^{+4.4}_{-5.4}$	&  $18.6^{+2.1}_{-2.3}$ &  12 \\
	&	& $43.29\pm0.10$ 	& $18.6^{+5.3}_{-1.6}$	&  $15.9^{+2.9}_{-2.5}$ &  12 \\
	&	& $43.01\pm0.11$ 	& $10.3^{+3.9}_{-4.0}$	&  $11.0^{+1.9}_{-2.0}$ & 12 \\
	&	& $43.26\pm0.10$ 	& $12.6^{+3.1}_{-4.0}$	&  $13.0^{+1.6}_{-1.4}$ & 12 \\
	&	& $43.32\pm0.10$ 	& $11.3^{+4.6}_{-3.4}$	&  $13.4^{+3.8}_{-4.3}$ & 12 \\
	&	& $43.46\pm0.09$ 	& $22.5^{+6.1}_{-5.2}$	&  $21.7\pm2.6$ &  12\\
	&	& $43.37\pm0.09$ 	& $14.3^{+1.0}_{-1.4}$	&  $16.4^{+1.2}_{-1.1}$ &  12\\
	&	& $43.18\pm0.10$ 	& $18.6^{+3.8}_{-4.2}$	&  $17.5^{+2.0}_{-1.6}$ & 12 \\
	&	& $43.52\pm0.09$ 	& $26.0^{+8.5}_{-3.7}$	&  $26.5^{+4.3}_{-2.2}$ &  12\\
	&	& $43.44\pm0.09$ 	& $25.0^{+8.8}_{-9.0}$	&  $24.8^{+3.2}_{-3.0}$ & 12 \\
	&	& $43.05\pm0.11$ 	& $13.0^{+13.0}_{-7.0}$	&  $6.5^{+5.7}_{-3.7}$ &  12\\
	&	& $43.05\pm0.11$ 	& $11.9^{+8.2}_{-10.9}$	&  $14.3^{+5.9}_{-7.3}$ & 12 \\

PG1426	&0.087	& $44.57\pm0.02$ 	& $82^{+66}_{-72}$	&  $95.0^{+29.9}_{-37.1}$ & 2 \\

Mrk\,817	&0.032	& $43.73\pm0.05$ 	& $9.2^{+2.9}_{-3.5}$	&  $14.04^{+3.41}_{-3.47}$ & 8\\

Mrk\,290	&0.030	& $43.11\pm0.06$ & $7.5^{+1.7}_{-1.8}$	&  $8.72^{+1.21}_{-1.02}$ & 8 \\

PG1613	&0.129	& $44.71\pm0.03$	& $72^{+47}_{-54}$	& $40.1^{+15.0}_{-15.2}$ & 2 \\

PG1617	&0.112	& $44.33\pm0.02$ 	& $31^{+32}_{-67}$	&  $71.5^{+29.6}_{-33.7}$ & 2 \\ 

PG1700	&0.292	& $45.53\pm0.01$ &  $124^{+328}_{-351}$	&  $251.8^{+45.9}_{-38.8}$ & 2 \\

3C\,390.3	&0.056	& $43.62\pm0.10$ 	& $23^{+22}_{-12}$	&  $23.6^{+6.2}_{-6.7}$ & 13 \\

NGC\,6814	&0.005	& $42.05\pm0.29$ 	& $6.9^{+1.1}_{-1.0}$	&  $6.64^{+0.87}_{-0.90}$ & 6 \\

Mrk\,509	&0.034	& $44.13\pm0.05$ & $95^{+19}_{-14}$	&  $79.60^{+6.1}_{-5.4}$ & 4 \\

PG2130	&0.063	& $44.14\pm0.03$ & $12.3^{+2.5}_{-1.9}$	&  $9.6\pm1.2$ & 1 \\

NGC\,7469	&0.016	& $43.56\pm0.10$ 	& $25.8^{+8.9}_{-10.4}$	&  $24.3\pm4.0$ & 14 \\
\tableline
\end{tabular}
\vskip 2pt
\parbox{3.4in}{ 
\small\baselineskip 9pt
\footnotesize
$^\dagger$Values are in the rest frame of the source. \\
{\bf Data references:} (1) \citealt{gr12}; (2) \citealt{kas00}; (3) \citealt{san97}; (4) \citealt{pet98}; (5) \citealt{den09}; (6) \citealt{ben09}; (7) \citealt{st94}; (8) \citealt{den09b} (9) \citealt{ben06}; (10) \citealt{den06}; (11) \citealt{san01}; (12) \citealt{pet02}; (13) \citealt{die98}; (14) \citealt{pet14}.
}

\end{center}
\end{table}

\subsection{Multi-epoch Data for NGC\,5548}
\label{sec5548}

NGC\,5548 has one of the best sets of light curves suitable for RM, covering more than a decade of high-cadence observations \citep{pet02}. Based on these data several works have shown that NGC\,5548 follows an intrinsic BLR-size-luminosity relation \citep{pet02,ben07,ben13}, which is consistent with the extrinsic (i.e., multi-object) one. Here  we analyze 13 consecutive sets of continuum and H$\beta$ light curves of NGC\,5548 spanning a year each (yearly definitions follow those of \citealt{pet02,ben13}), and benchmark the performance of the optimized estimators defined in \S2 by quantifying the scatter around the BLR-size--luminosity relation. Host-subtracted luminosity estimates for each epoch were taken from \citet[and see \S \ref{secBentz}]{ben13}. 

Figure \ref{N5548} shows versions of the intrinsic BLR-size--luminosity relations as obtained by each of the estimators defined in \S2. The relative performance of the estimators is insensitive to whether weighted or unweighted versions are used. We find that the VN estimator outperforms the other measures of randomness, and leads to a tighter correlation between time-delay and luminosity with $\rho \sim 0.7$ (time delays are quoted in Table \ref{T1}). Conversely, the cumulative sums estimator is the worst performer (leading in some cases to negative lags and to  $\rho <0.5$). These findings confirm our simulations (Fig. \ref{2b}). 

Given the advantages of the VN scheme over the other estimators explored here, we henceforth focus on its performance, and  consider its weighted version with nontrivial implementations of $G$- and $W$-implementations unless otherwise specified.

\subsection{The Palomar-Green Quasar Sample of \citet{kas00}}
\label{secPG}

Here we benchmark the performance of the VN schemes using the spectrophotometric data of \citet{kas00} for 17 Palomar-Green (PG) quasars, and compare it to the default implementation of the ZDCF (see also \citealt{kas00}). Time delays reported for the latter are obtained using the maximum likelihood approach of \citet{al13} and the corresponding correlation functions are identical to those reported in \citet[see their Fig. 4]{kas00} and hence are not shown.  

Time-delay measurements for the H$\beta$ line are reported here for two implementations of the VN algorithm: including nontrivial weights and flags (leading to the $\tau_{\rm VN}$ column in Table \ref{T2}), and for an unweighted version operating over the full extent of $F$ (i.e.,  setting $G=W=1$ and leading to the $\tau_{\rm VN}^{G=W=1}$ column in Table \ref{T2}). Below we comment on the statistical properties of the lags obtained with further details provided in Appendix C. That said, the reader is advised not to over-interpret the statistics given the sample size and data quality.

We find the results of the weighted and unweighted VN schemes to be similar with $\rho >0.9$, and that their time delays are both comparably correlated with the quasar luminosity and have a similar $\rho$-value to that of the ICCF scheme \citep[and our Table \ref{T2}]{kas00}. In contrast, the ZDCF time delays lead to a significantly weaker correlation with luminosity. Further, while the VN scheme leads to positive delays in all cases, the ZDCF reports negative delays in $\lesssim 20$\% of the cases. This hints at the superior stability of the VN scheme, as also indicated by our simulations, and perhaps reflects also on the overestimated lag uncertainties obtained by the FR/RSS scheme; see Appendix B. We also find that the VN results are very well correlated with the ICCF time delays \citep[with $\rho >0.96$; see Table \ref{T2} and][]{kas00}. Taken together, our findings suggest that the VN scheme has two promising features: (1) it has the major advantage of being largely model- and binning-independent while (2) producing more stable results than default implementations of discrete cross-correlation function schemes.

\begin{table*}
\caption{Von Neumann Analysis of the Sample of Palomar-Green Quasars of \citet{kas00}}
\label{T2}
\vspace{-0.15in}
\begin{center}
\begin{tabular}{lllllll}
\tableline
	    & 	    	& $L_{\rm opt}$	& $\tau_{\rm Kaspi}$	& $\tau_{\rm ZDCF}$	& $\tau_{\rm VN}$ & $\tau^{\rm G=W=1}_{\rm VN}$	\\
object  &  $z$	& 	 $(10^{44}\,{\rm erg~s^{-1}})$ 		& (days$^\P$) 		& (days$^\P$)		& (days$^\P$)  & (days$^\P$)\\
\tableline
PG0026+129	&0.142	& $10.3\pm1.5$	& $125^{+29}_{-36}$ 	& $142\pm29$	& $90^{+37}_{-27}$  & $98^{+12}_{-44}$\\
PG0052+251	&0.155	& $9.1\pm1.6$	& $99^{+30}_{-31}$ 	& $111^{+54}_{-51}$	&$79\pm23$  & $77\pm31$\\
PG0804+761	&0.100	& $8.6\pm 1.6$	& $151^{+26}_{-24}$ 	& $86^{+94}_{-24}$	& $135^{+27}_{-40}$ & $126^{+12}_{-9}$\\
PG0844+349	&0.064	& $2.21\pm 0.23$ & $13^{+14}_{-11}$	& $-19^{+28}_{-27}$	&  $4^{+30}_{-25}$ &  $22^{+26}_{-10}$\\
PG0953+414	&0.239	& $16.6\pm 2.2$ & $187^{+27}_{-33}$	& $211^{+51}_{-108}$	&  $191^{+80}_{-36}$ & $191^{+27}_{-22}$ \\
PG1211+143	&0.085	& $5.57\pm 0.90$ & $103^{+32}_{-44}$	& $-6^{+158}_{-22}$	&  $11^{+13}_{-15}$ & $142^{+128}_{-103}$ \\
PG1226+023	&0.158	& $91.1\pm11.1$ & $382^{+117}_{-96}$	& $253^{+274}_{-88}$	&  $393^{+147}_{-140}$ & $379^{+120}_{-90}$\\
PG1229+204	&0.064	& $1.21\pm0.13$ & $36^{+32}_{-18}$	& $61^{+154}_{-42}$	&  $32^{+31}_{-18}$ & $56^{+41}_{-5}$ \\
PG1307+085	&0.155	& $7.54\pm0.76$ & $108^{+46}_{-115}$	& $22^{+227}_{-124}$	&  $103^{+83}_{-86}$ & $140^{+125}_{-108}$ \\
PG1351+640$^\dagger$	&0.087	& $4.95\pm0.49$ & $247^{+162}_{-78}$	& $268^{+40}_{-70}$	&  $202^{+200}_{-135}$ & $333^{+139}_{-319}$ \\
PG1411+442	&0.089	& $4.22\pm0.36$ & $118^{+72}_{-71}$	& $62^{+64}_{-33}$	&  $107^{+86}_{-138}$ & $166^{+133}_{-144}$ \\
PG1426+015	&0.086	& $5.21\pm0.80$ & $115^{+49}_{-68}$	& $39^{+106}_{-119}$	&  $89^{+72}_{-78}$ & $121^{+78}_{-77}$\\
PG1613+658	&0.129	& $9.5\pm1.2$	& $44^{+20}_{-23}$ 	& $86\pm35$	& $81^{+53}_{-611}$ & $129^{+78}_{-77}$\\
PG1617+175	&0.114	& $3.00\pm0.52$ & $78^{+30}_{-41}$	& $31^{+117}_{-260}$	&  $34^{+36}_{-75}$ & $70^{+68}_{-78}$\\ 
PG1700+518	&0.292	& $42.3\pm2.9$ & $114^{+246}_{-235}$	& $-122^{+332}_{-10}$	&  $160^{+424}_{-454}$ & $124^{+328}_{-351}$\\
PG1704+608	&0.371	& $56.0\pm8.2$ & $437^{+252}_{-391}$	& $269^{+323}_{-73}$	&  $390^{+246}_{-237}$ &  $508^{+248}_{-355}$ \\
PG2130+099	&0.061	& $2.85\pm0.26$ & $188^{+136}_{-27}$	& $204^{+52}_{-100}$	&  $181^{+118}_{-83}$ &  $191^{+89}_{-45}$\\
\tableline
\multicolumn{3}{l}{Pearson's correlation coefficients$^\flat$:} & \multicolumn{3}{c}{$\llcorner  \_ \_ \_ \_ \_ \_ \_ \_ \_ \_$ 0.96 $ \_ \_ \_ \_ \_  \_ \_ \_ \_ \_ \lrcorner$} \\
\multicolumn{3}{l}{} & \multicolumn{2}{c}{$\llcorner  \_\_ \_ \_ \_ 0.76   \_ \_ \_ \_   \lrcorner$} \\
\multicolumn{5}{l}{} & \multicolumn{2}{c}{$\llcorner \_ \_ \_  \_ 0.92  \_ \_ \_ \_   \lrcorner$} \\
\multicolumn{4}{l}{} & \multicolumn{2}{c}{$\llcorner \_ \_ \_  \_ 0.71  \_ \_ \_ \_   \lrcorner$} \\
\multicolumn{2}{l}{} & \multicolumn{5}{c}{$\llcorner  \_ \_ \_  \_ \_  \_ \_  \_ \_ \_ \_ \_ \_ \_  \_  \_  \_  \_  \_  \_  \_  \_  \_  \_  \_ \_ 0.59^\ddag  \_  \_  \_  \_ \_  \_  \_  \_  \_ \_  \_ \_  \_  \_  \_   \_ \_ \_ \_   \_ \_  \_ \_ \_ \_ \_ \lrcorner$} \\

\multicolumn{2}{l}{} & \multicolumn{4}{c}{$\llcorner  \_ \_  \_ \_ \_ \_ \_ \_ \_  \_  \_  \_  \_  \_  \_  \_  \_  \_  \_ 0.63^\ddag  \_  \_  \_  \_ \_  \_  \_  \_  \_ \_  \_ \_  \_  \_  \_   \_ \_ \_ \_ \lrcorner$} \\
\multicolumn{2}{l}{} & \multicolumn{3}{c}{$\llcorner  \_  \_  \_  \_  \_  \_  \_  \_  \_  \_  \_ <0.48^\ddag   \_  \_  \_ \_  \_  \_  \_  \_ \_  \_ \_  \lrcorner$} \\
\multicolumn{2}{l}{} & \multicolumn{2}{c}{$\llcorner \_ \_  \_  \_ \_ 0.61^\ddag  \_ \_  \_  \_  \_ \lrcorner$} & \\

\tableline
\end{tabular}
\vskip 2pt
\parbox{7.0in}{ 
\small\baselineskip 9pt
\footnotesize
$^\P$Values are in the observed frame. \\
$^\dagger$H$\alpha$ lag measurements were considered as in \citet{kas00}.\\
$^\ddag$After correcting for cosmological time dilation and using logarithmic values (negative time delays were ignored). \\
$^\flat$Correlation coefficients should not be overinterpreted because of the limited sample size and the uncertainties associated with individual measurements. Uncertainties on the correlation coefficients are not provided for similar reasons.
}
\end{center}
\end{table*}

\subsection{The Sample of Active Galactic Nuclei of \citet{ben13}}
\label{secBentz}

\begin{figure*}
\epsscale{1.17}
\plottwo{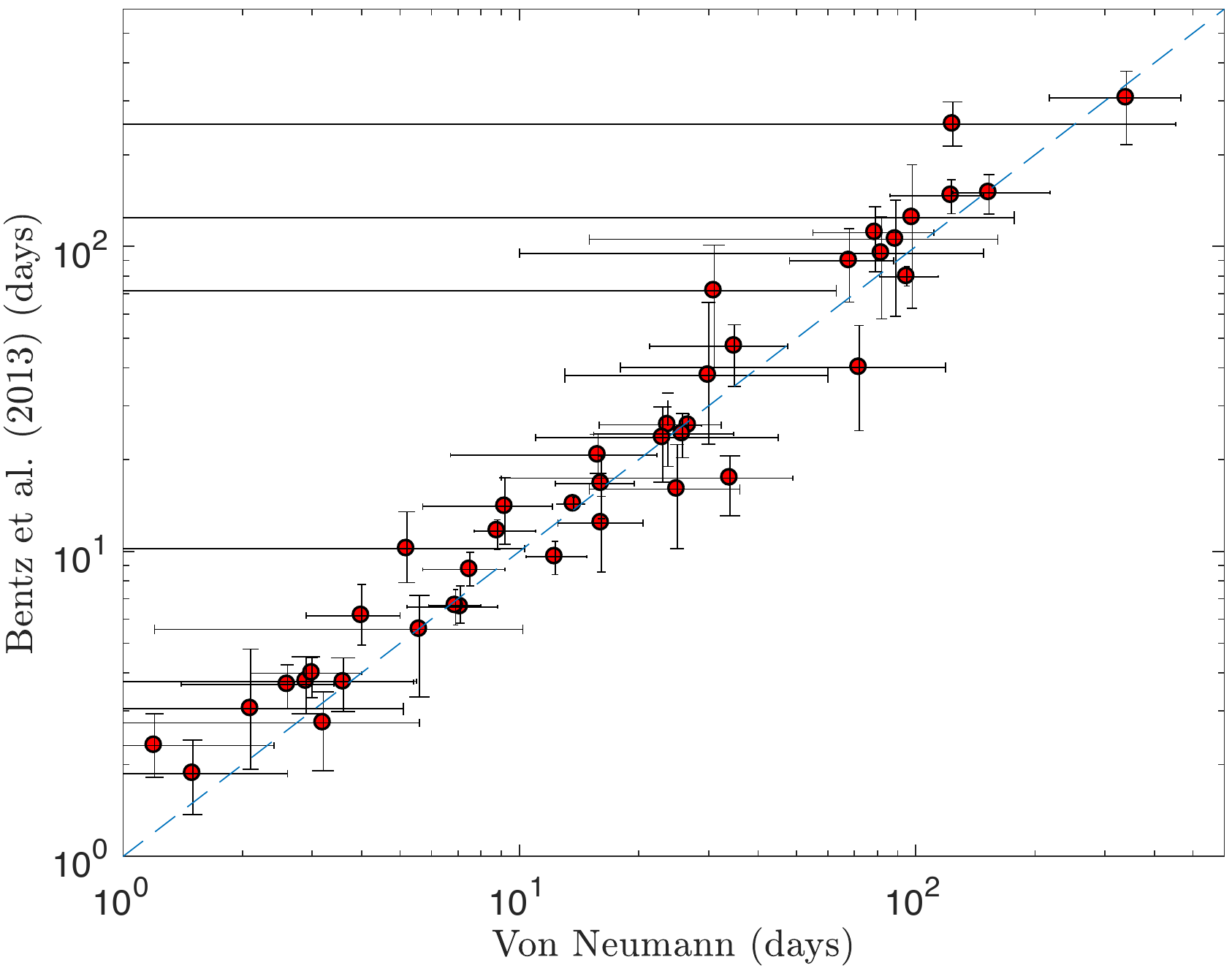}{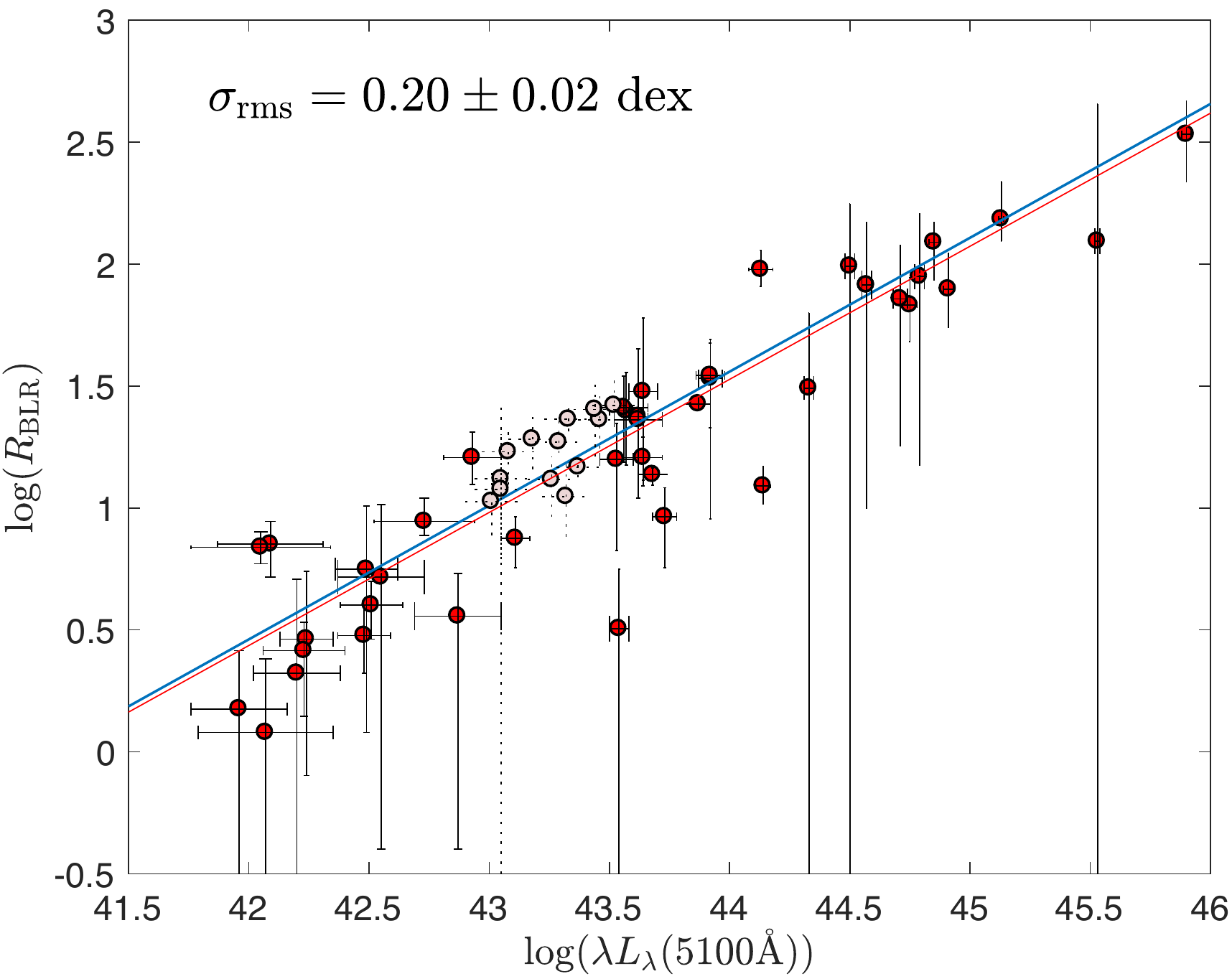}
\caption{Application of the optimized and weighted VN scheme for the sample of \citet{ben13}. {\it Left:} Comparison of the time delays obtained here to those reported in \citet[see references therein]{ben13}, which are largely based on the ICCF scheme (see Table \ref{T1}). The agreement between the two sets of measurements is good with $\rho \simeq 0.94$. The error bars reported here appear to be overestimated by $\sim 40$\% (see text). {\it Right:} The BLR-size--luminosity relation obtained using the VN scheme. We find consistency with the results of \citet{ben13} for the targets analyzed in Table \ref{T1} (red points) including multi-epoch time-lag measurements of NGC\,5548 (Fig. \ref{N5548}). The scatter in the points around the best-fit relation of \citet{ben13}  (blue line) is $\simeq 0.2$\,dex, and hence is comparable to that reported by \citet{ben13} and \citet{zu11}. The unweighted power-law fit to our measurements (red line) is consistent with the fit of \citet{ben13} (see text).}
\label{LR}
\end{figure*}

Here we revisit the sample of active galactic nuclei (AGN) of \citet{ben13}, for which the nuclear luminosity has been estimated by subtracting off the host's contribution to the aperture flux, resulting in a tighter BLR size--luminosity relation. In particular, we reanalyze the continuum and H$\beta$ data for all sources listed in Table \ref{T1} using the optimized and weighted VN scheme with nontrivial $G$-factor implementations. More information about individual targets may be found in Appendix C.

Time-lag measurements are in most cases robust, with the delays being highly significant, as expected given the overall data quality of the sample. That said, the values reported at the short-lag end of the distribution appear to be somewhat less significant than the values reported by \citet{ben13}, which may relate to the overall overestimation of the FR/RSS uncertainties when applied to the VN scheme (see below and Appendix B). Overall, a statistical comparison of the time delays obtained here to those of \citet[and references therein]{ben13} shows good agreement with $\rho \gtrsim 0.9$ (Fig. \ref{LR}). 

The expected correspondence between our results and the lags reported in \citet{ben13} could be used to estimate the degree to which the quoted lag uncertainties are reasonable. To this end, we take a simplified approach and symmetrize the lower and upper measurement uncertainties reported in Table \ref{T1}, and consider the ratio $\tau_{\rm VN}/\tau_{\rm Bentz}$ and its uncertainty with errors added in quadrature. With the expectation of a ratio of unity, we calculate the reduced $\chi^2$ and find its value to be $<0.5$. Reducing the lag uncertainties by $\simeq 45$\% leads to a reduced $\chi^2\simeq 1$, and in agreement with our calculations in Appendix B (note that this is merely a global estimate and may not reflect on the reliability of the measurement uncertainties for individual sources, nor on the reliability of the quoted uncertainties for the ICCF scheme). 

Our version of the BLR-size-luminosity is shown in Figure \ref{LR} together with multi-epoch results for NGC\,5548 of \S\ref{sec5548}. The results are consistent with the best-fit relation of \citet{ben13} of the form ${\rm log}(R_{\rm BLR}/1\,{\rm day})=K+\alpha\,{\rm log}(L_{\rm opt}/10^{44}\,{\rm erg~s^{-1}})$, and an unweighted fit to the von Neumann results gives a power-law slope $\alpha \simeq 0.546$ and an intercept $K\simeq 1.527$, which are consistent with the values reported in \citet{ben13} of $K=1.527\pm0.031$ and $\alpha=0.533^{+0.035}_{-0.033}$.

We find the dispersion in our lag measurements relative to the (theoretical) BLR-size-luminosity relation\footnote{After excluding Mrk\,142 and NGC\,3227 for reasons given in \citet{ben13}. Quoted uncertainties on $\sigma_{\rm rms}$ were obtained by applying a random subset selection (RSS) scheme -- akin to that discussed in \citet{pet98} -- to the data shown in the right panel of Figure \ref{LR} and quoted in Table \ref{T1}. These do not include lag uncertainties contribution to the scatter due to reasons discussed in Appendix B.} to be $\sigma_{\rm rms}=0.20\pm0.02$\,dex, which is comparable to the values found in \citet{zu11} and \citet{ben13}; note, however, the differences in the samples (objects and epochs) as well as estimates of source luminosity considered in each work.

\section{Summary}

In this paper we seek to expand the range of methods available for time-lag determination in the field of quasar reverberation-mapping. To this end we introduce a class of statistical estimators, which was originally used to measure the degree of randomness or complexity of a dataset (e.g., in cryptographic applications), and adapt it to RM. 

Six different randomness/complexity measures are considered, among which the von Neumann and Lempel--Ziv estimators (the latter is at the core of many lossless data compression algorithms)  and appropriate extrema criteria for time-lag measurements are defined for each. At least two distinct implementations are considered per estimator (e.g., error-weighted and unweighted schemes) with the feature common to all being the minimal set of assumptions used to analyze the data. Specifically, none of the estimators considered here requires polynomial interpolation schemes to be defined, nor do they employ binning in correlation space. Further, they do not rely on ergodicity arguments and (tunable) stochastic models for quasar variability, hence time-delay measurements are restricted to the information embedded in the light curves being processed.

Among the various measures considered in this work, we find that an optimized implementation of the VN estimator \citep[see][for a similar but not identical approach]{pel94} performs best with the added benefit of being mathematically simple and computationally efficient to evaluate. We benchmark its performance using a large set of simulated light curves, and study the effects of data sampling rate, time-series duration, S/N, light-curve spikes, and correlated errors on the deduced lags. The effects of the properties of the quasar -- the power density spectrum of its light curve and the BLR transfer function -- on the deduced time-delay statistics are also investigated. We find that the optimized VN outperforms current implementations of the discrete correlation scheme for much of the parameter space relevant to RM, and leads to better agreement with the true lags.

Applying the optimized VN estimator to existing data of varying quality, we find its performance to be on par with that of other (more model-dependent) time-lag measurement schemes. In particular, the parameterization of the BLR-size--luminosity relation of \citet{ben13} is recovered, and the scatter around this theoretical relation is found to be comparable to that reported in other studies.  

It is found that the optimized VN estimator can, under favorable conditions, be used in conjunction with existing methods, such as the discrete correlation function scheme, to obtain refined lags and to identify biases or the breakdown of the assumptions underlying other methods of RM. Such refinement and verification of time delays is crucial for alleviating systematics and for reducing the scatter in the size-luminosity relation, as well as for identifying departures from it \citep[e.g.,][]{du16}, with implications for quasar physics and the use of quasars as standard cosmological candles.

To conclude, we find that measures of randomness significantly add to our arsenal of tools for RM, and that further exploration of the vast literature in the field is warranted. In particular, some of the schemes outlined here may be further refined, and additional measures of randomness/complexity not explored here may achieve superior performance as far as quasar RM is concerned.

\acknowledgements 

This research has been supported in part by grants 950/15 and 848/16 from the Israeli Science Foundation (ISF), and by a Deutsche Forschungsgemeinschaft (DFG) grant HA3555/14-1. Computations were carried out on the Hive high-performance computer cluster at the University of Haifa, which is partly funded by ISF grant 2155/15. D.\,C. thanks Shai Kaspi and Uri Keshet for fruitful discussions that motivated this work, Boris Chorny and Keith Horne for wise advice, and Edi Barkai for invaluable and persistent support. We thank the referee for illuminating feedback.

\newpage

\appendix

\section{A scheme for computing the optimized von Neumann estimator}

Here we outline an efficient computing scheme\footnote{See \url{http://chelouche.haifa.ac.il/codes/VonNeumann.m} for a Matlab\textsuperscript{\textregistered} version of the full code, and \url{http://www.pozonunez.de/astro_codes/python/vnrm.py} or \url{http://www.pozonunez.de/astro_codes/idl/vnrm.pro} for partial Python/IDL\textsuperscript{\textregistered} implementations.} for determining the global minimum of the VN estimator given the renormalization of $F_2^\tau$ and $F_1$ as defined in equations \ref{f1} and \ref{f2}. We first note that we can write
\begin{equation}
F'=\left \{  \left ( t_i, f_i +\alpha_i(\eta f_i +\epsilon)  \right ) \right \}_{i=1}^N,~~~F''=\left \{  \left ( t_i, f_i -(1-\alpha_i)(\eta f_i +\epsilon)  \right ) \right \}_{i=1}^N,
\end{equation}
where $\alpha_i=1$ if the $i$th data point originates from $F_2^\tau$, and $\alpha_i=0$ otherwise. Using these, we can write equation \ref{dd0} as
\begin{equation}
\mathfrak{T}=\frac{1}{2}   \sum_{i=1}^{N-1} \left \{ \left [ f_i+\alpha_i (\eta f_i + \epsilon) -   f_{i+1} -\alpha_{i+1} (\eta f_{i+1} + \epsilon )  \right ]^2 + 
\left [ f_i+(\alpha_i -1) (\eta f_i + \epsilon) -   f_{i+1} -(\alpha_{i+1} -1) (\eta f_{i+1} + \epsilon )  \right ]^2
 \right \} .
 \label{dd}
\end{equation}
Requiring a minimum for $\mathfrak{T}$, at any given $\tau$, we obtain the following set of two linear equations, ${\bm {\mathit y}}(\tau) = {\bm {\mathit A}}(\tau) {\bm {\mathit x}} $, where
\begin{equation}
{\bm x} = \left (
\begin{array}{c}
\displaystyle \epsilon \\
\displaystyle \eta \\
\end{array}
\right ) ,~~{\bm y} = \left (
\begin{array}{c}
\displaystyle 2\sum_{i=1}^{N-1} (\alpha_i -\alpha_{i+1})(f_{i+1}-f_i) \\
\displaystyle \sum_{i=1}^{N-1} [(2\alpha_i-1)f_i-(2\alpha_{i+1}-1)f_{i+1}](f_{i+1}-f_i)  \\
\end{array}
\right ) ,
\label{opt1}
\end{equation}
and
\begin{equation}
{\bm {\mathit A}} = \left (
\begin{array}{cc}
\displaystyle 2\sum_{i=1}^{N-1} (\alpha_i -\alpha_{i+1})^2 & \displaystyle \sum_{i=1}^{N-1} (\alpha_i-\alpha_{i+1} ) [ (2\alpha_i -1)f_i - (2\alpha_{i+1}-1) f_{i+1}] \\
\displaystyle \sum_{i=1}^{N-1} (\alpha_i-\alpha_{i+1} ) [ (2\alpha_i -1)f_i - (2\alpha_{i+1}-1) f_{i+1}] & \displaystyle \sum_{i=1}^{N-1} [ (f_i-f_{i+1})^2+2(\alpha_i-\alpha_{i+1})^2f_if_{i+1}]  \\
\end{array}
\right ) .
\label{opt2}
\end{equation}
The solutions for $\eta(\tau)$ and $\epsilon(\tau)$, inserted back into equation \ref{dd}, yield the minimal $\mathfrak{T}(\tau)$ (see, e.g., Figs. \ref{PG}, \ref{AGN}), from which $\tau_0^\ast$ may be deduced at a modest computational cost. A similar, but not identical approach may reduce the computational load of the optimization scheme also for the other estimators discussed in this paper but is not further explored here. 

The inclusion of weights, $W$, and flags, $G$ (see \S \ref{flags}) is straightforward and amounts to multiplying Eq. \ref{dd} by $W_{i,i+1}G_{i,i+1}$ while maintaining proper normalization (Eq. \ref{vnn}). The derivation of the optimized schemes follows, as outlined above.

\section{The FR/RSS scheme and the von Neumann estimator}

The FR/RSS method of \citet[which includes refinements of the scheme of \citealt{pet98} due to \citealt{wel99}]{pet04} is used here to estimate the uncertainties on the time delay between the light curves of individual sources. While this method has been shown to provide reasonable, albeit somewhat conservative, lag uncertainties for ICCF schemes, it is yet to be shown to provide reasonable values also for the optimized VN scheme. To this end, we simulated the light curves for $\sim 10^5$ sources (\S3), and carried out calculations of uncertainty in lag for each using the FR/RSS scheme with $10^4$ realizations. We find that, when averaged over the entire mock population, the input lag falls outside the 16\%-84\% range obtained via the FR/RSS scheme in $\lesssim 10\%$ of the cases (for the model parameterization explored here, and assuming random cadence), suggesting that the uncertainties are conservative. Requiring that the input delay falls within the designated uncertainty interval in $\simeq 68$\% of the cases (within one standard deviation around the peak for normal distributions) requires that the quoted FR/RSS uncertainties be reduced by $\gtrsim 40\%$ on average. A comparative analysis of the results of \citet{ben13} and the time delays obtained here supports this conclusion (\S\ref{secBentz}). Nevertheless, these results may not hold for individual targets, or for all object parameterizations and and light-curve cadences, and we therefore quote the standard FR/RSS uncertainties in Table 1. A more complete treatment of the estimation of the time-lag uncertainty for the VN RM scheme is beyond the scope of the present paper.

\section{Notes on Individual Sources}

\begin{figure*}
\epsscale{1.17}
\plotone{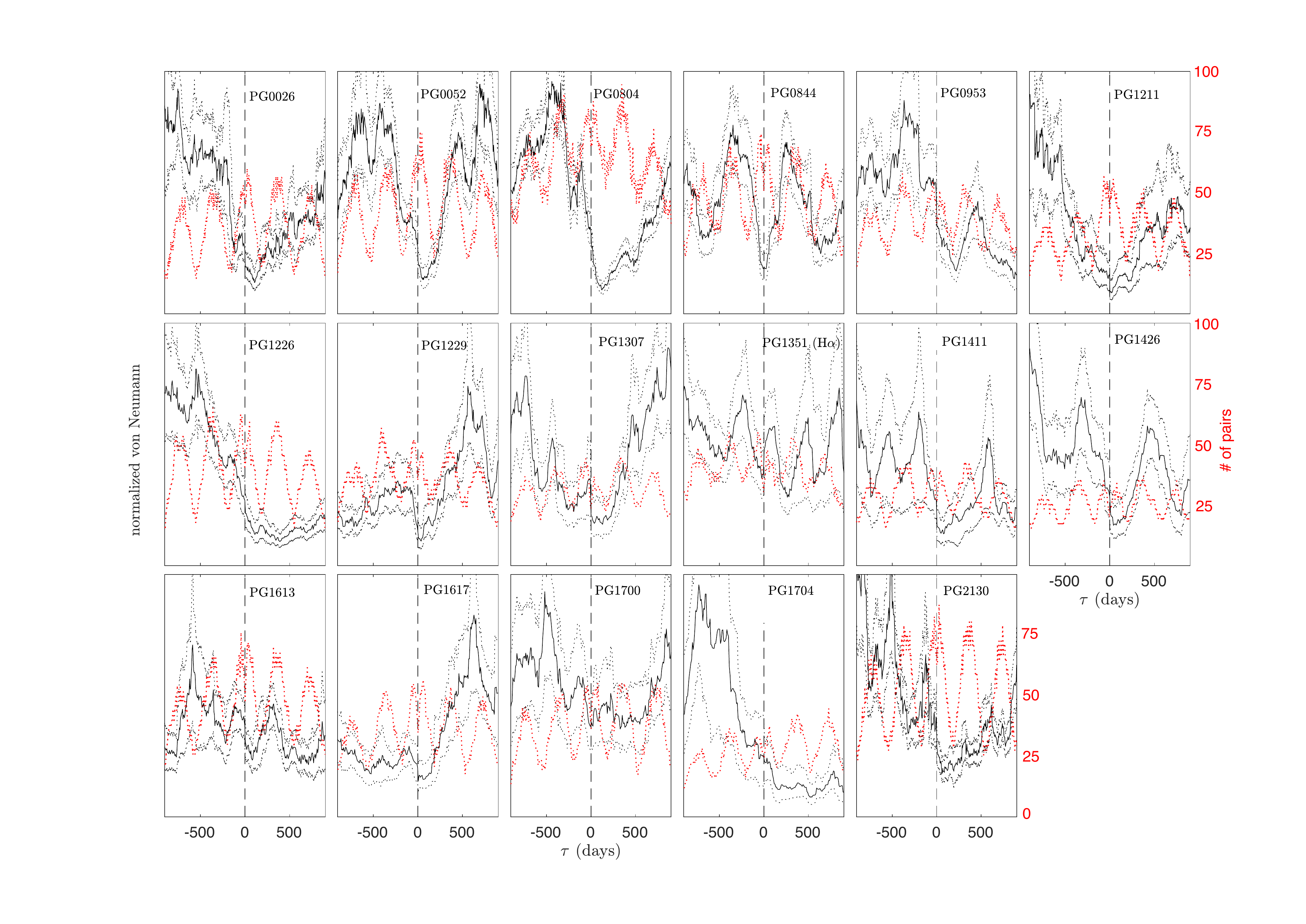}
\caption{Results for the H$\beta$ emission line based on the optimized VN estimator for the PG sample of quasars of \citet{kas00}. Curves show the optimized VN mean (solid black line) and the 16\%, 84\%  envelopes around it (dotted black lines) as obtained by the FR/RSS scheme. The time shift, $\tau$, at which a $\mathfrak{T}$-minimum occurs corresponds to the time delay (values for each source are provided in Table 1). This figure may be compared to Figure 4 in \citet{kas00} showing ZDCF and ICCF results for each of the targets, with qualitative agreement implied. Results for the H$\alpha$ line are shown for PG\,1351 for reasons discussed in \citet{kas00}. Red curves show the number of pairs of points in the VN sum with $G=1$ (see Eq. \ref{gflag}).}
\label{PG}
\end{figure*}

\begin{figure*}
\epsscale{1.17}
\plotone{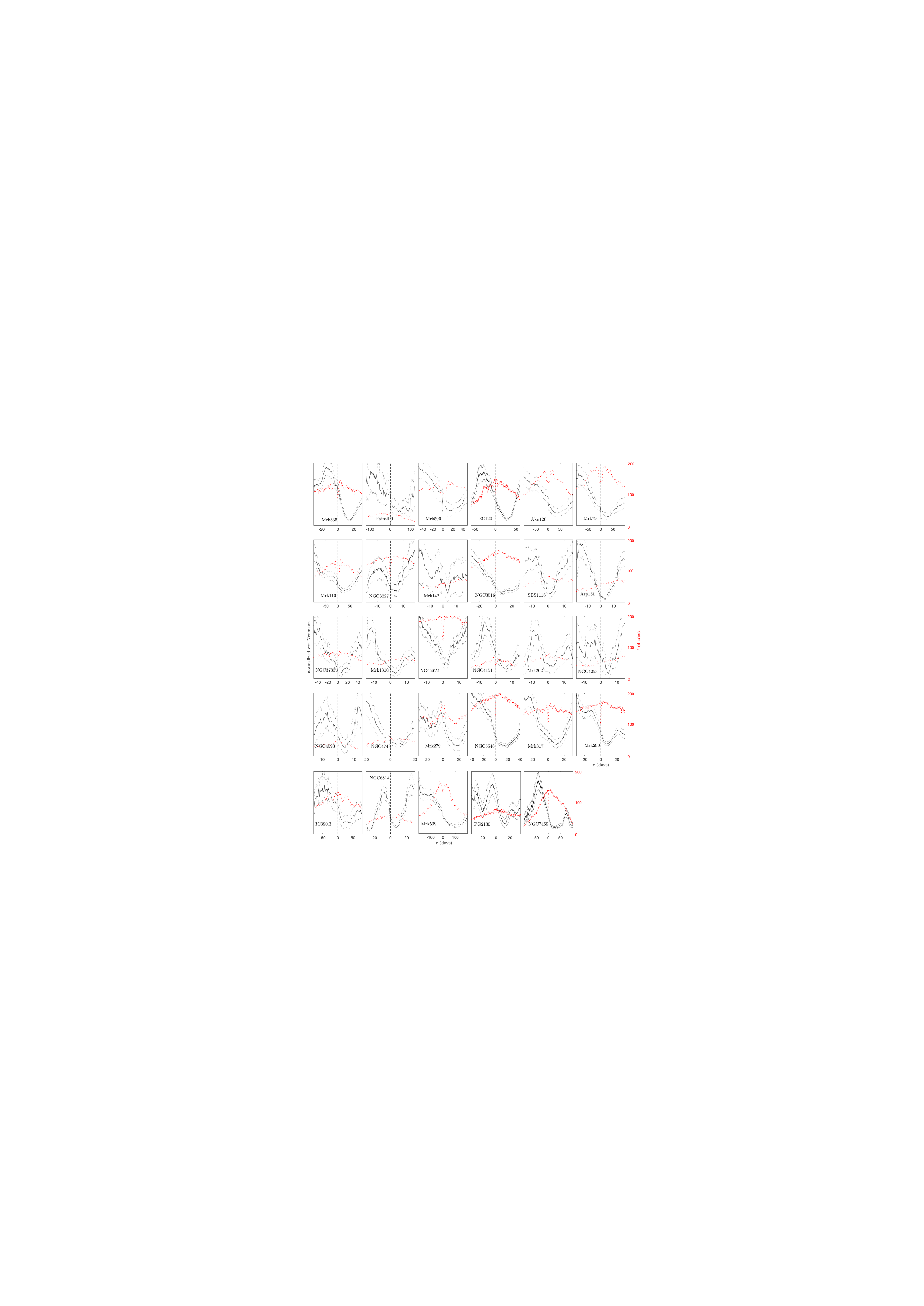}
\caption{Same as Fig. \ref{PG} for the remaining objects in the complilation of AGN of \citet{ben13}. Values for the time delays for each of the sources are provided in Table 2.}
\label{AGN}
\end{figure*}

Results for the optimized (weighted) VN estimator are provided for each of the sources in Tables \ref{T1} and \ref{T2} with the estimator values and uncertainty contours shown in Figures \ref{PG} and \ref{AGN}. Note that the superior quality of the light curves that characterize objects in the compilation of \citet[and references therein]{ben13} results in more regular VN estimators with well defined troughs compared to the sample of \citet{kas00}. This also reflects on the width of the error "snakes" around the mean VN values (dotted black curves in Figs. \ref{PG} ,\ref{AGN}), and results in better defined time delays. Note that the values of the VN estimator at different values of $\tau$ are correlated, as are their uncertainties. Also shown in Figures \ref{PG} and \ref{AGN} in red are the numbers of pairs of points with $G=1$ (Eq. \ref{gflag}) that determine $\mathfrak{T}(\tau)$. The effect of discarding zero-lag pairs of points is evident in all sources, as are the seasonal gaps that might affect some time-delay measurements for the PG quasars \citep[see also][]{zu11} but are less relevant for most AGNs in the compilation of \citet{ben13}. 

Below we comment on anomalies encountered while analyzing the sources and epochs considered here. 

\begin{itemize}
\item
{\it Mrk\,79}:  A second minimum at 200 days is ignored and the search for time delays is restricted to the range $[-120,120]$\,days. Using the default search window of $[-T/3,T/3]$ we get $153^{+123}_{-50}$\,days.
\item
{\it Mrk\,110}: The lag quoted in Table \ref{T1} was obtained by analyzing the full light curve for this source, while the values quoted from \citet{ben13} are based on the mean values reported there and the uncertainty on their standard deviation (i.e., quoted lag uncertainties from \citet{ben13} do not include the FR/RSS contribution). 
\item
{\it PG\,0953}: The VN time delay is stable so long as the search window is $<600$\,days\,$\simeq T/4$.
\item
{\it NGC\,3227}: The VN time delay is stable so long as the search window is  narrower than $[-20,20]$\,days\,$\simeq T/4$.
\item
{\it NGC\,4151}: The time-delay search interval was constrained to  the interval $[-15,15]$\,days since there exists a second VN minimum at 20\,days whose inclusion leads to a solution of $7.7^{+2.5}_{-9.2}$\,days.
\item
{\it Mrk\,202}: Time delay search interval was restricted to the interval $[-13,13]$\,days\,$\simeq T/4$.
\item
{\it PG\,1307}: We take an approach akin to \citet{kas00}, and find a stable solution within a search interval with a lower bound of $-160$\,days (the upper bound matters little to the end result).
\item
{\it PG\,1229}: We restrict the search window to the range $[-500,500]$\,days, because a second VN minimum exists at $<-500$\,days, which is reminiscent of the negative peak seen in the ZDCF for this source (see Fig. 4 in \citealt{kas00}).
\item
{\it Mrk\,279}: We search for lags within the range $[-50,50]$\,days range since there exists another VN minimum at $\gtrsim 50$\,days resulting in $67^{+51}_{-33}$\,days.
\item
{\it NGC\,5548}: The results shown in Figure \ref{AGN} were obtained for the light curves of \citet{den09}. The analysis of individual epochs based on 13 years of data from \citet{pet02} is not shown.
\item
{\it PG\,1411}: The time-delay search window is restricted to $[-500,500]$\,days since a second VN minimum exists outside this range.
\item
{\it PG\,1426}: The time-delay search window is restricted to $[-500,500]$\,days since a second VN minimum exists outside this range.
\item
{\it PG\,1613}: The time-delay search window is restricted to $[-300,300]$\,days since a second VN minimum exists outside this range.
\item
{\it 3C\,390.3}: We find that different implementations of the VN algorithm (e.g., weighted vs. unweighted) change the structure of the VN trough in a non-negligible way, which can exceed the formal uncertainties provided by the FR/RSS scheme. Nevertheless, the delay is positive and $<50$\,days for all implementations. Better data are required to better constrain the lag in this source.
\item
{\it NGC\,6814}: A local VN minimum exists also at negative delays where the number of pairs contributing to the signal is  reduced (see Fig. \ref{AGN}). We therefore restrict the search window to $>-20$\,days\,$\simeq-T/3.4$.
\item
{\it PG\,2130}: Two very different delays are reported for this source: Table \ref{T2} and Figure \ref{PG} quote delays for the data set of \citet{kas00} while Table \ref{T1} and Figure \ref{AGN} quote lags for the \citet{gr12} dataset. For more information about potential sampling issues for this source see \citet{gr08,gr12}.
\end{itemize}

\end{document}